\documentclass[]{aa}
\usepackage{txfonts}
\usepackage{graphics,graphicx,epstopdf,epsfig,url}
\usepackage{relsize}
\usepackage{dcolumn}
\usepackage{bm}
\usepackage[normalem]{ulem}
\usepackage{times}
\usepackage{natbib}
\usepackage{color}
\usepackage{multirow}
\usepackage{hyperref}
\usepackage{braket}
\usepackage[utf8]{inputenc}
\usepackage{xcolor}
\usepackage{hypcap}
\usepackage{tensor}
\hypersetup{
    colorlinks=true, 
    citecolor=blue, 
    filecolor=black, 
    linkcolor=blue,
    urlcolor= black
}
\def\codename{\texttt{SCALAR} }
\def\codenamex{\texttt{SCALAR}}

\begin{document}

\title{Solitons in the dark: non-linear structure formation with fuzzy dark matter}

  \author{Mattia Mina\inst{1}
          \and
          David F. Mota\inst{1}
          \and
          Hans A. Winther\inst{1}
          }

\institute{Institute of Theoretical Astrophysics, University of Oslo, 0315 Oslo, Norway\\
\email{mattia.mina@astro.uio.no}
}

\date{Date}

\abstract{
We present the results of a full cosmological simulation with the new code \codenamex, where dark matter is in form of fuzzy dark matter, described by a light scalar field with a mass of $m_{\rm B} = 2.5 \times 10^{-22}$ eV and evolving according to the Schr\"{o}dinger-Poisson system of equations. 
In comoving units, the simulation volume is $2.5 ~ h^{-1} {\rm Mpc}$ on a side, with a resolution of $20~h^{-1}{\rm pc}$ at the finest refinement level. 
We analyse the formation and the evolution of central solitonic cores, which are found to leave their imprints on dark matter density profiles, resulting in shallower central densities, and on rotation curves, producing an additional circular velocity peak at small radii from the center. We find that the suppression of structures due to the quantum nature of the scalar field results in an shallower halo mass function in the low-mass end compared to the case of a $\Lambda$CDM simulation, in which dark matter is expected to cluster at all mass scales even if evolved with the same initial conditions used for fuzzy dark matter. Furthermore, we verify the scaling relations characterising the solution to the Schr\"{o}dinger--Poisson system, for both isolated and merging halos, and we find that they are preserved by merging processes. We characterise each fuzzy dark matter halo in terms of the dimensionless quantity $\Xi \propto \left | E_{\rm halo} \right |/M_{\rm halo}^3$ and we show that the core mass is tightly linked to the halo mass by the core--halo mass relation $M_{\rm core}/M_{\rm halo} \propto \Xi^{1/3}$. We also show that the core surface density of the simulated fuzzy dark matter halos does not follow the scaling with the core radius as observed for dwarf galaxies, representing a big challenge for the fuzzy dark matter model as the sole explanation of core formation.

}


\maketitle

\section{Introduction}

The dynamics of the large-scale structures that are observed today is well captured by the standard model of cosmology, the $\Lambda$CDM model, where our Universe is described in terms of ordinary matter and radiation ($\sim 5$\%), cold dark matter ($\sim 25$\%), and dark energy in form of a cosmological constant ($\sim 70$\%) \citep{2018arXiv180706209P}. 
The formation and the evolution of structures in our Universe, seeded by small density inhomogeneities, has found its main driver in the dark matter component. This form of cold and pressureless medium heavily dominates the matter content of the Universe, accounting for roughly $\sim 85$\% of the total non-relativistic matter component.
The fluid description of cold dark matter (CDM) accurately describes the properties of the observed Universe at large scales and the $\Lambda$CDM model is well constrained down to galactic scales. 
At galactic and subgalactic scales, the formation and evolution of structures is highly non-linear and it can mainly be investigated by means of numerical simulations. Since the first studies of this kind, cosmological {\it N}-body simulations \citep{aarseth_2003,2002AA...385..337T,2005MNRAS.364.1105S} have improved in many different aspects.
Initially \citep{1964ApNr....9..313A}, {\it N}-body simulations could only model the dynamics of a small group of particles interacting by nothing else than gravity, thus describing the CDM component of the Universe.
Nowadays, pure cosmological {\it N}-body simulations have evolved into hydrodynamic simulations \citep{2014Natur.509..177V,2014MNRAS.444.1453D,2015MNRAS.446..521S}, including not only the baryonic physics, but also a large variety of astrophysical phenomena, which are needful to simulate realistic galaxies.
However, the $\Lambda$CDM model struggles at reproducing some basic small-scale properties of the observed Universe and, at galactic and subgalactic scales, it is facing a big challenge.

% Cusp-core 
Cosmological dark matter only simulations have found a universal density profile describing dark matter halos of any mass and size. Density profiles of individual CDM halos peak in their innermost region and they are well fitted by a cuspy Navarro-Frenk-White (NFW) profile \citep{1996ApJ...462..563N}:
\begin{align}\label{eq:nfw}
\rho_{\rm NFW}(r) = \rho_0\left [\frac{r}{R_s} \left ( 1 + \frac{r}{R_s} \right )^2\right]^{-1} ~ ,
\end{align}
where $\rho_0$ corresponds to the central density of the dark halo. The scale radius $R_s$ marks the transition point between a log-slope of $\gamma \sim -1$ in the central part and a log-slope of $\gamma \sim -3$ in the outskirt of the dark halo.
However, many observations of rotation curves in dwarf galaxies have shown a preferred cored isothermal profile \citep{1995ApJ...447L..25B}, with nearly constant density within its core radius \citep{1994Natur.370..629M,1994ApJ...427L...1F,2004MNRAS.351..903G,10.1111/j.1365-2966.2009.15004.x,2010AdAst2010E...5D}.

% Missing satellite
In addition, due to its cold and collisionless nature, dark matter clumps exist at all scales. Dark halos are populated by substructures, also known as subhalos, of any size. This prediction was soon verified by the the first numerical simulations, but the count of the number of substructures far exceeds the number of subhalos found in the Local Volume \citep{1999ApJ...522...82K,1999ApJ...524L..19M}. The number of subhalos of a galaxy such as the Milky Way is roughly one order of magnitude smaller than predicted by the $\Lambda$CDM model.

%TBTF
Furthermore, several studies have shown that the process of star formation is very sensitive to astrophysical and environmental processes and, in low-mass halos, it can be very stochastic.
However, many substructures predicted by the $\Lambda$CDM model are simply too massive to have failed star formation, representing an additional source of discrepancy \citep{2011MNRAS.415L..40B,10.1111/j.1365-2966.2012.20695.x}. 

% Scaling relations
Moreover, in spite of the wide diversity of structures found in the Universe, astronomical observations suggest a tight connection between dynamical and baryonic properties of galaxies. These scaling relations still represent a big challenge for the $\Lambda$CDM model at small scales. An example is the baryonic Tully-Fisher relation (BTFR), linking the total baryonic mass of a galaxy with its asymptotic circular velocity \citep{2000ApJ...533L..99M}. While within the $\Lambda$CDM model the latter scales as $V_{\rm circ}^3 \propto M_b$ with the baryonic mass, the observed trend suggests a scaling of $V_{\rm circ}^4 \propto M_b$. 
In general, despite the enormous improvement of numerical techniques and astrophysical models, it is still not trivial to reproduce all scaling relations at once. 

Nowadays, it is still unclear where discrepancies between theoretical predictions of numerical simulations and astronomical observations at small scales have their origin.
On one hand, dark matter could not be as cold and collisionless as previously thought, but its true nature remains unknown. On the other hand, small-scale baryonic astrophysical processes could very well resolve these tensions \citep{10.1111/j.1365-2966.2012.20696.x,2013ApJ...765...22B,2017MNRAS.472.4786G}, but they have an extremely complex dynamics and hydrodynamic simulations might not capture all the relevant physics involved in small-scale astrophysical phenomena.

At the present day, weakly interactive massive particles (WIMPs) are still considered, by many, to be one of the most likely dark matter candidates. Without any success so far, many ongoing experiments are trying to detect such particles, but the unprobed region of the parameter space is closing down on the neutrino floor. Beyond the neutrino floor, it would be impossible to observe any signature left by WIMPs, as the signal would be drowned in the solar neutrino background.

This motivates the search for alternative dark matter candidates. Nowadays, models involving ultra-light scalars, such as ultra-light axions (ULAs) and fuzzy dark matter (FDM), are among the most promising alternatives to WIMPs \citep[see e.g.][]{1983PhLB..120..137D,1983PhLB..120..127P,1996PhRvD..53.2236L,Peebles_2000,2000PhRvL..85.1158H,2016PhR...643....1M,2017PhRvD..95d3541H}. 
This class of models are very appealing and they gained a lot of attention in the past decade. Not only they predict distinct and observable signatures at small scales, but they can alleviate tensions between theoretical predictions of the $\Lambda$CDM model and astronomical observations.
The dynamics of FDM and ULAs is described by the classical Schr\"odinger equation. Recently, an exiguous number of numerical tools where developed to study the dynamics of this class of models and the first numerical simulations were performed.

Traditionally, there are two main categories of numerical algorithms that have been developed for this purpose.
The Madelung formulation of quantum mechanics \citep{1926NW.....14.1004M} defines a system of hydrodynamic equations, where the quantum nature of a collection of extremely light bosons is encoded in a pressure-like term, called quantum pressure. On scales below the bosons de Broglie wavelength, quantum pressure prevents the collapse of dark matter particles by counteracting gravity.
The Madelung equations can be used to model the physics of ULAs and FDM in particle-based fluid simulations. In this case, different numerical schemes have been designed in order to discretise the quantum pressure, and the Madelung equations are solved by means of traditional smoothed particle hydrodynamics (SPH), as suggested by \citet{2015PhRvD..91l3520M}. 
For example, by using a particle-based approach, \citet{2018ApJ...853...51Z} and \citet{2018arXiv180108144N} successfully reproduced the expected density profiles of dark halos.
In spite of being faster than other algorithms implemented in grid-based codes, the hydrodynamic approach can suffer of a lack of accuracy. In fact, the Madelung formulation is known to break down in extremely low density regions, such as voids and interference nodes, as the quantum pressure can develop singularities \citep{2014PhRvD..90b3517U}.

Alternatively, the dynamics of the ULAs and FDM can be described by solving the Schr\"odinger-Poisson system in a grid-based approach. In this case, the wave-function is discretised on a grid and the non-linear Schr\"odinger equation can be solved using different techniques.
For example, several grid-based codes implement spectral methods and they solve the governing equations in Fourier space \citep{2009ApJ...697..850W,2017MNRAS.471.4559M,2018JCAP...10..027E,Mocz_2020}. 
Other grid-based codes employ a Taylor method to discretise the time evolution operator, in order to compute the formal solution of the non-linear Schr\"odinger equation, \citep{2014NatPh..10..496S,2019PhRvD..99f3509L,mina2019scalar}.
The general evolution of a Universe with FDM was initially investigated in \citet{2009ApJ...697..850W}, while the first high resolution cosmological simulation with FDM was performed in \citet{2014NatPh..10..496S}, where they studied the behaviour of the FDM fluid within dark halos. In \citet{2017MNRAS.471.4559M} and \citet{Mocz_2020}, instead, they simulated the formation and the evolution galaxies and dark halos with the Bose-Einstein condensate dark matter (BECDM) model.

The Schr\"odinger-Poisson system has also been introduced as an alternative tool to sample the six dimensional phase-space of a system of collisionless self-gravitating particles \citep{1993ApJ...416L..71W}. In this regard, \citet{2017PhRvD..96l3532K} and \citet{2018PhRvD..97h3519M} showed that it is possible to recover the classical behaviour in the limit of $\hbar \to 0$, with excellent agreement with the solution of the standard Vlasov equations.
Although grid-based numerical methods are very accurate, they are also slower than particle-based codes. For this reason, in some of these works, only two-dimensional applications were considered, as full three-dimensional applications were too expensive in terms of computational resources.

Despite being in qualitative agreement, different numerical studies have reached different quantitative conclusions. In particular, the core and the halo masses are expected to be tightly linked by the core--halo mass relation, but the scaling between the two is found to be different in almost every numerical study of structure formation involving ULAs and FDM \citep{2014NatPh..10..496S,Schive_2014,2016PhRvD..94d3513S,2017MNRAS.471.4559M,2018PhRvD..97h3519M}.
Such differences can be attributed to many factors, including numerical algorithms.
A exhaustive benchmark study of different numerical tools is still missing, but it is important to test and verify the dynamics of ultra-light scalar fields with different codes in order to reach a quantitative agreement on the theoretical predictions of this class of models.

In this paper we present a high resolution cosmological simulation of a Universe where the entire budget of dark matter is in the form of FDM. The simulation is performed by using the \codename (Simulation Code for ultrA Light Axions in \texttt{RAMSES}) code. 
The purpose of this work is to investigate and characterise the dynamics of dark matter halos within the paradigm of FDM, together with formation and evolution of their central solitonic cores, which are a key signature of models based on ultra-light scalar fields. The study of how such cores form and evolve is essential for testing the model against observations. 
For example, a soliton core is found to fit very well the cores observed in some dwarf galaxies, which in turn have been used to constrain, or even determine, the mass of the light boson assuming that the formation mechanism of such cores is solely due to the phenomenology of ULAs or FDM \citep{2014NatPh..10..496S,2015MNRAS.451.2479M,2016MNRAS.460.4397C,2017MNRAS.472.1346G,2015MNRAS.450..209B}.
On the other hand, \citet{2020arXiv200611111B} recently argued that dark matter cores previously predicted by cosmological simulations involving this class of models are inconsistent with observations if FDM dark halos form following the classical hierarchical paradigm.

The structure of the paper is as follows: in Section~\ref{sec:theory} we describe the theory behind the class of dark matter models involving light scalar fields, in Section~\ref{sec:simulation} we introduce the simulation and we briefly describe the numerical methods used to solve the Schr\"odinger-Poisson system. Then, we present the results of the simulation in Section~\ref{sec:results}, before concluding in Section~\ref{sec:conc}.

\section{Theory}\label{sec:theory}
In this section we summarise the phenomenology and the dynamics of light scalar fields in a general context, without introducing any specific model motivated by particle physics. Then, starting from the phenomenology of a light scalar field, we describe them in cosmological context, we derive the governing equations, and we briefly describe the their dynamics. The equations are given in natural units, where $c=\hbar=1$.

\subsection{Scalar fields as Dark Matter}
A complex scalar field has an internal global $U(1)$ symmetry, which is spontaneously broken when it acquires a vacuum expectation value. Thus the two component of the complex scalar field are reconfigured in a massive mode and a Goldstone boson. The Goldstone boson is a massless angular degree of freedom, which is invariant under shift transformations. 
However, at some energy scale, non-perturbative physics becomes relevant and it explicitly breaks the shift symmetry, leading to a preferred field configuration and, thus, a potential for the Goldstone boson.
The potential must respect the residual discrete shift symmetry, since the Goldstone boson still represents an angular degree of freedom, and it must therefore be periodic.
By denoting with $\phi$ the Goldstone boson, with $f_{\phi}$ the energy scale of the spontaneous symmetry breaking, and with $\Lambda_{\phi}$ the energy scale at which non-perturbative effects becomes relevant, the potential can be generally written as $V(\phi) = \Lambda_{\phi}^4 U(\phi/f_{\phi})$, where $U(\phi/f_{\phi})$ is periodic.
Although the explicit form of the potential depends on the underlying model, one of its simplest forms is:
\begin{align}
V(\phi) = \Lambda_{\phi}^4 \left [ 1-\cos \left (\frac{\phi}{f_{\phi}} \right ) \right ] ~ .
\end{align}
In order to study the dynamics of $\phi$ in a model independent way, we only consider small displacements of the field from the minimum of the potential. Thus, the potential can be expressed as a Taylor series. The leading term of the Taylor expansion is the mass term and the potential can be approximated as follows:
\begin{align}
V(\phi) \sim \frac{1}{2} m_{\rm B}^2 \phi^2 ~ ,
\end{align}
where $m_{\rm B} = \Lambda_{\phi}^4/f_{\phi}^2$ corresponds to the mass of the boson. Higher order contributions to the Taylor expansion of the potential would include terms describing self-interactions and interactions with other standard model fields. However, those terms are suppressed by higher powers of $f_{\phi}$ and we do not consider them, as they are not relevant for this work.
Typically, the parameter $f_{\phi}$ lies in between the traditional energy scale of the grand unification theory $E_{\rm GUT} \sim 10^{16}$ GeV and the Plank energy $E_{\rm Pl} \sim 10^{18}$ GeV.
Again, the energy scale of non-perturbative physics is extremely sensitive to the details of the underlying model and it is not relevant for this work. 

The only relevant assumption in the context of dark matter cosmology revolves around the boson mass. 
Typical FDM models consider a scalar field with a mass in the range of $10^{-24} < m_{\rm B} < 10^{-22} ~\rm{eV}$, which is of particular interest for current observations \citep[see e.g][]{2010PhRvD..82j3528M,2015MNRAS.451.2479M,2016MNRAS.460.4397C,2017MNRAS.472.1346G,2019JCAP...07..045B}.
In fact, due to their extremely small mass, FDM particles manifest their wave nature on astronomical scales. While large-scale predictions would essentially be the same as for the $\Lambda$CDM model, the quantum nature of the dark matter fluid would suppress the formation of structure at small scales, providing a natural solution to the small-scale problems of the $\Lambda$CDM model \citep{2015MNRAS.451.2479M,2017PhRvD..95d3541H}.
Indeed, the corresponding de Broglie wavelength of FDM particles:
\begin{align}\label{eq:dB}
\lambda_{dB} = \frac{\lambda}{2\pi} = \frac{1}{m_{\rm B} v} ~ ,
\end{align}
is much bigger than the mean inter-particle separation. Under this circumstances, FDM particles have high ground state occupation number and they form a Bose-Einstein condensate (BEC). As a consequence, the system behaves as a macroscopic quantum state and it can be described by a single wave-function evolving according to the classical Schr\"odinger equation.

\subsection{Fuzzy Dark Matter}
In general relativity (GR), a spin-0 real scalar field minimally coupled with the metric is described by the action:
\begin{align}
S_{\phi} = \int \rm{d}^4 x \sqrt{-g} \left [ \frac{1}{2} \tensor{g}{^\mu ^\nu} \partial_{\mu}\phi \partial_{\nu} \phi - V(\phi) \right ] ~ .
\end{align}
This action is only valid after the symmetry is spontaneously broken and once non-perturbative effects have switched on. 
The equation of motion of the field can be obtained by varying the action with respect to the field itself, and it is in the form of a Klein-Gordon equation:
\begin{align}\label{eq:klein-gordon}
\frac{1}{\sqrt{-g}} \partial_{\mu} \left [ \sqrt{-g} ~ \tensor{g}{^\mu ^\nu} \partial_{\nu} \phi\right]  - \frac{\partial V}{\partial \phi} = 0 ~ .
\end{align}
The corresponding energy-momentum tensor can be derived, instead, by varying the action with respect to the metric:
\begin{align}\label{eq:en-mom}
\tensor{T}{^\mu _\nu} = \tensor{g}{^\mu^\alpha} \tensor{\partial}{_\alpha} \phi \tensor{\partial}{_\nu} \phi - \frac{\tensor{\delta}{^\mu_\nu}}{2} \left [ \tensor{g}{^\alpha^\beta} \tensor{\partial}{_\alpha}\phi \tensor{\partial}{_\beta}\phi + V(\phi) \right ].
\end{align}

\paragraph{Background evolution.}
The background evolution of the field can be studied under the assumption of homogeneity and isotropy. For this purpose, by computing the d'Alembert operator for the Friedman-Robertson-Walker (FRW) metric and by replacing the potential with its Taylor expansion up to the leading order, the equation of motion of the field reduces to:
\begin{align}\label{eq:kg-oscillator}
\frac{d^2 \phi}{dt} + 3 H \frac{d\phi}{dt} + m_{\rm B}^2 \phi = 0 ~ ,
\end{align}
where $H$ denotes the Hubble expansion rate. This equation corresponds to a simple harmonic oscillator, with a time dependent friction term determined by the underlying background cosmology. 
The damping ratio of the system can be defined as:
\begin{align}
\zeta = \frac{3H}{2\omega_0} ~ ,
\end{align}
where $\omega_0 = m_{\rm B}$ corresponds to the natural frequency of the system. 
The behaviour of the system can be characterised by two regimes. At early stages, the mass term is completely negligible compared to the Hubble term. Thus, in this regime, we can disregard the mass term and Eq.~\eqref{eq:kg-oscillator} describes an overdumped harmonic oscillator, as $\zeta \gg 1$. In this case, the value of the field remains frozen to its initial value. 
However, as the Hubble rate drops as $H\sim t^{-1}$, the damping ratio decreases and, at $t = t_{\rm osc}$, the condition $\zeta = 1$ defines the crossover between the overdumped regime and a new one, where the field starts to coherently oscillate. In the limit of $\zeta \ll 1$, we can neglect the friction term in Eq.~\eqref{eq:kg-oscillator} and the system is described by an underdumped harmonic oscillator.
In particular, when the Universe is matter dominated, the scale factor evolves as $a \propto t^p$ and the exact solution of Eq. \eqref{eq:kg-oscillator} reads:
\begin{align}\label{eq:kg-oscillator-sol}
\phi(a) = a^{-3/2} (t/t_{\rm ini})^{1/2} \left [ C_1 J_n(m_{\rm B}t) +C_2 Y_n(m_{\rm B}t) \right ] ~ ,
\end{align}
where $n= (3p-1)/2$, $J_n(x)$ and $Y_n(x)$ are Bessel function of the first and the second kind respectively, and $t_{\rm ini}$ is the initial time.
In order to better understand the behaviour of the field, we can define its background energy density and pressure as follows:
\begin{align}
\rho_{\rm B} &= \frac{1}{2} \left [ \left ( \frac{d\phi}{dt} \right )^2 + m_{\rm B}^2 \phi^2 \right ] \label{eq:bg-dens} ~ , \\
p_{\rm B} &= \frac{1}{2} \left [ \left ( \frac{d\phi}{dt} \right )^2 - m_{\rm B}^2 \phi^2 \right ] \label{eq:bg-pres} ~ .
\end{align}
In the overdumped regime, the first term between parenthesis in both Eq.~\eqref{eq:bg-dens} and Eq.~\eqref{eq:bg-pres} is negligible and the field have an effective equation of state $w_{\rm B} = p_{\rm B}/\rho_{\rm B} = -1$. As a consequence, at early stages, the field effectively behaves as a dark energy component.
On the other hand, by plugging Eq.~\eqref{eq:kg-oscillator-sol} into Eq.~\eqref{eq:bg-dens} and Eq.~\eqref{eq:bg-pres}, it is possible to show that, in the underdamped regime, the equation of state is the same as any non-relativistic component (e.g. CDM) and the background energy density evolves as $\rho \propto a^{-3}$.
As long as the crossover $\zeta = 1$ occurs before the matter-radiation equality, this kind of models are suitable as alternative dark matter models. 
In the $\Lambda$CDM model, the Hubble rate at matter-radiation equality is roughly $H(a_{\rm eq}) \sim 10^{28}$ eV. This poses a lower bound to the boson mass: models involving a scalar field with a mass heavier than $\sim 10^{28}$ eV represents a potential dark matter candidate.

As a consequence, assuming that the transition between the two regimes takes place within the radiation dominated epoch, the present day relic abundance of FDM can be written as:
\begin{align}
\Omega_{\rm FDM} = 0.1 \left ( \frac{m_{\rm B}}{10^{-22}~\rm{eV}} \right)^{1/2} \left ( \frac{f_{\phi}}{10^{17} ~ \rm{ GeV}} \right )^2~.
\end{align}

\paragraph{Non-linear dynamics.}
In the context of cosmological structure formation, we focus on the underdamped regime and we assume throughout the rest of the paper that the whole dark matter budget of the Universe is in form of FDM. 
In order to study growth of perturbations in a FDM Universe, we consider a perturbed FRW metric instead, in the Newtonian gauge. Thus, in the week field limit, the metric tensor is given by:
\begin{align}\label{eq:metric}
ds^2 = -\left ( 1+2\Phi \right ) dt^2 + a^2(t) \left( 1-2\Phi \right) d\mathbf{r}^2~,
\end{align}
where $\Phi$ denotes the Newtonian gravitational potential.
Plugging the non-zero components of the metric tensor in the Klein-Gordon equation, Eq.~\eqref{eq:klein-gordon}, leads to:
\begin{align}
\left ( 1-2\Phi \right ) \frac{\partial^2\phi}{\partial t^2} &+3H \left ( \frac{\partial \Phi}{\partial t} -2\Phi +1 \right ) \frac{\partial\phi}{\partial t} \nonumber \\ 
&- \left ( 1+2\Phi \right ) \frac{1}{a^2} \nabla^2 \phi - m_{\rm B}^2 \phi = 0.
\end{align}
Instead, plugging the non-zero components of the metric tensor in the "00" component of the energy-momentum tensor, Eq.~\eqref{eq:en-mom}, gives the energy density of the field:
\begin{align}\label{eq:en-dens}
\rho_{\rm B} = \frac{1}{2} \left [ \left (1+2\Phi \right ) \frac{\partial\phi}{\partial t} + m_{\rm B}\phi^2 + \frac{\left (1-2\Phi \right )}{a^2} \partial^i \phi \partial_i \phi \right ].
\end{align}
However, since rapid temporal fluctuations of the density of the field do not contribute at all to the gravitational potential, we can explicitly disregard the high frequency part of the spectrum by considering the non-relativistic limit. For this purpose, we express $\phi$ in terms of a complex scalar field $\psi$:
\begin{align}\label{eq:f_cut}
\phi = \frac{1}{\sqrt{2 m_{\rm B}}} \left ( \psi e^{-im_{\rm B}t} + \psi^{*} e^{im_{\rm B}t} \right ) ~ ,
\end{align}
thus filtering out the contribution of high frequency modes, i.e. $\omega \sim m_{\rm B}$.
As a consequence, we can safely assume that $|\ddot{\psi}| \ll m_{\rm B} |\dot{\psi} |$ and the equation of motion for the complex scalar field $\psi$ reduces to:
\begin{align}\label{eq:schrodinger}
i \left ( \frac{\partial \psi}{\partial t} + \frac{3}{2} H \psi \right ) = \left ( -\frac{1}{2m_{\rm B}^2a^2} \nabla^2 + m_{\rm B} \Phi \right ) \psi ~ .
\end{align}
Eq.~\eqref{eq:schrodinger} corresponds to a non-linear Schr\"odinger equation, generalised to the case of an expanding Universe. 
Furthermore, in the non-relativistic limit, the leading term in the energy density of the field, Eq.~\eqref{eq:en-dens}, corresponds to:
\begin{align}
\rho_{\rm B} = m_{\rm B}\left | \psi \right |^2.
\end{align}
In order to study the non-linear clustering of dark matter in a cosmological context, Eq.~\eqref{eq:schrodinger} is coupled to the Poisson equation which, under the previous assumptions, reads:
\begin{align}\label{eq:poisson}
\nabla^2 \Phi = 4\pi G a^2 (\rho_{\rm B}-\overline{\rho}_{\rm B}) ~ ,
\end{align}
and it describes how the gravitational potential reacts to fluctuations in the density field.

It is also possible to recast the Schr\"odinger equation in a system of hydrodynamic equations. In this case, we express the wave-function $\psi$ in polar coordinates:
\begin{align}
\psi = \sqrt{\frac{\rho_{\rm B}}{m_{\rm B}}} e^{ i\theta} ~ ,
\end{align}
and we describe the behaviour of the dark matter fluid in terms of the macroscopic quantities:
\begin{align}
\rho &= m_{\rm B} \left | \psi \right |^2 ~, \\
\mathbf{v} &= \frac{\nabla \theta}{m_{\rm B}} ~.
\end{align}
By replacing the previous definitions into the Schr\"odinger equation and considering separately the real and the imaginary parts, the dynamics of the macroscopic fluid follows the system of equations:
\begin{align}
& \frac{\partial\rho}{\partial t} +3H\rho +\frac{1}{a} \nabla \cdot \left (\rho \mathbf{v} \right ) = 0 \label{eq:Mad1} ~ ,\\
& \frac{\partial \mathbf{v}}{\partial t} + H\mathbf{v} + \frac{1}{a} \left ( \mathbf{v} \cdot \nabla \right ) \mathbf{v}  = \frac{1}{m_{\rm B}a}\nabla \left (\Phi + Q \right ) \label{eq:Mad2} ~ .
\end{align}
Here, the term $Q$ is the so-called quantum potential and it is defined as:
\begin{align}\label{eq:qp}
Q = -\frac{1}{2m_{\rm B}a^2} \frac{\nabla^2\sqrt{\rho}}{\sqrt{\rho}} ~ .
\end{align}
Eq.~\eqref{eq:Mad1} and Eq.~\eqref{eq:Mad2} correspond to the Madelung formulation of quantum mechanics. Density and velocity assume now a classical meaning. While equation Eq.~\eqref{eq:Mad1} is the same as the classical continuity equation and it describes the conservation of mass, the Euler-like equation Eq.~\eqref{eq:Mad2} expresses the conservation of momentum. Contrary to its classical counterpart, Eq.~\eqref{eq:Mad2} does not have a classical pressure term. However, the term $Q$ is equivalent to a pressure term, generating a certain "stiffness" in the field, which in turn resists the compression due to gravity. The quantum pressure only acts in certain regions of the Universe. In particular, when the density field tends to zero, the term $Q$ vanishes. As a consequence, when the quantum potential is absent or negligible, Eqs.~\eqref{eq:Mad1}-\eqref{eq:Mad2} describe a system of particle only interacting by means of gravity and they have the same form as CDM fluid equations. 
In a cosmological context, the quantum pressure is expected to be important in extremely high density regions, such as the innermost part of a dark matter halo, and on scales of the de Broglie wavelength of dark matter particles, but everywhere else the fields behave as the classical CDM.

\section{Simulation}\label{sec:simulation}
In this section we summarise the numerical setup we adopt for our simulation and the numerical schemes used by \codenamex, in order to solve the governing equations of FDM.

\subsection{Numerical setup}
For this work, we simulate a Universe where the whole dark matter budget is in the form of FDM. For this purpose, we employ the new adaptive mesh refinement (AMR) code \codename to evolve a scalar field with a mass of $m_{\rm B} =2.5 \times 10^{-22}$ eV, in a cosmology with present day matter and dark energy density parameters of respectively $\Omega_{\rm FDM} = 0.3$ and $\Omega_{\Lambda}=0.7$, a dimensionless Hubble constant of $h=0.67$ and a linear power spectrum normalisation of $\sigma_8 = 0.8$. 
The simulation volume of $2.5 ~h^{-1} {\rm Mpc}$ on a side is discretised with $512^3$ cells at the domain grid, and with up to eight refinement levels. 
Thus, while the resolution of the domain grid is only $\Delta x \sim 5 ~h^{-1}{\rm kpc}$, the effective resolution of our simulation is approximately $\Delta x \sim 20 ~h^{-1}  {\rm pc}$. 
According to Eq.~\eqref{eq:dB}, if we consider the typical circular velocity profile in the innermost part of an average dark matter halo, quantum effects are expected appear on scales of roughly $1~h^{-1}{\rm kpc}$.
Thus, the extremely high resolution is enough to capture the behaviour of the field in regions dominated by quantum effects.
The initial linear power spectrum is computed with the publicly available code\footnote{The code is availiable at https://github.com/dgrin1/axionCAMB} \texttt{AxionCAMB} \citep{2015PhRvD..91j3512H} and it is used to construct the initial conditions for the scalar field at redshift $z=200$ (please refer to Appendix \ref{sect:initial_conditions} for how initial conditions are generated). The Universe is then evolved until redshift $z=2.5$. 
The same initial conditions are used to simulate a $\Lambda$CDM Universe with the same cosmology, by using the \texttt{RAMSES} code.

\subsection{Numerical schemes}
In \codenamex, the solution of the non-linear Schr\"odinger equation is discretised on an AMR grid, where finer resolutions are only employed in regions where features of the wave-function are more demanding. 
Provided the wave-function at time $t_n$, the formal solution of the non-linear Schr\"odinger equation at $t_{n+1} = t_n +\Delta t$ reads:
\begin{align}
\psi \left ( \mathbf{x}, t_{n+1} \right ) = U \left ( t_{n+1}, t_n \right ) \psi \left ( \mathbf{x}, t_{n} \right ) ~ .
\end{align}
The propagator $U\left ( t_{n+1}, t_n \right )$, also known as the time evolution operator, links the wave-function at different times and it is discretised as follows:
\begin{align}
U \left ( t_{n+1}, t_n \right ) = \exp \left [ -i \hat{H} \left ( \mathbf{x}, t_n \right ) \Delta t \right ] ~ ,
\end{align}
where $\hat{H} \left ( \mathbf{x}, t_n \right )$ denotes the Hamiltonian of the system.
The Lie-Trotter formula is used to split kinetic and potential terms in the Hamiltonian, which are denoted by $\hat{K}\left ( \mathbf{x}, t_n \right )$ and $\hat{W}\left ( \mathbf{x}, t_n \right )$, respectively. Thus, the solution of the non-linear Schr\"odinger equation at time $t_{n+1}$ is computed as:
\begin{align}
\psi \left ( \mathbf{x}, t_{n+1} \right ) = \exp \left [ -i \hat{W} \left ( \mathbf{x}, t_n \right ) \Delta t \right ] \exp \left [ -i \hat{K} \left ( \mathbf{x}, t_n \right ) \Delta t \right ]  \psi \left ( \mathbf{x}, t_{n} \right ) ~ .
\end{align}
First, the term involving the free kinetic part of the Hamiltonian is expanded in Taylor series and it is applied to the wave-function as follows:
\begin{align}\label{eq:taylor-kin}
\bar{\psi} \left ( \mathbf{x}, t_{n+1} \right ) =  \left [ 1 + \left ( \frac{i \Delta t}{2 m_{\rm B}} \nabla^2 \right ) + \frac{1}{2} \left ( \frac{i \Delta t}{2 m_{\rm B}} \nabla^2 \right )^2 + \dots \right ]  \psi \left ( \mathbf{x}, t_{n} \right )~.
\end{align}
Here, the laplacian is computed by using the standard second-order finite difference formula. Furthermore, in Eq.\eqref{eq:taylor-kin}, \codename only considers terms up to $\mathcal{O}(dt^3)$.
Then, the phase rotation induced by the potential is calculated and the wave-function at time $t_{n+1}$ is computed as:
\begin{align}
\psi \left ( \mathbf{x}, t_{n+1} \right ) = \exp \left [ -i m_{\rm B} \Phi \left ( \mathbf{x}, t_n \right ) \Delta t \right  ] \bar{\psi} \left ( \mathbf{x}, t_{n+1} \right )~.
\end{align}
In order to ensure good conservation properties, \codename can employ a secondary solver. For this purpose, density currents $\mathbf{j}$ between AMR cells are computed at cell interfaces and at half time-step. Then, the associated continuity equation is solved on top of the non-linear Schr\"odinger equation as:
\begin{align}
\rho \left (\mathbf{x},t_{n+1} \right ) = \rho \left (\mathbf{x},t_{n} \right ) - \frac{\Delta t}{\Delta x} \left [ \mathbf{j} \left (\mathbf{x}_{i+1/2},t_{n+1/2} \right ) - \mathbf{j} \left (\mathbf{x}_{i-1/2},t_{n+1/2} \right ) \right ]~.
\end{align}
The new dark matter density is used to rescale the wave-function which was previously advanced by the Schr\"odinger solver. This process ensures a good level of mass conservation, which is needed for cosmological simulations.

\noindent
The algorithm used by \codename to solve the Poisson equation, instead, is the same as the one originally implemented in the \texttt{RAMSES} code. 
At the domain level, the gravitational potential is computed in Fourier space, by using a spectral solver. At finer levels of the AMR hierarchy, \codename switches to its multi-grid solver, which determines the solution of the Poisson equation by using a successive over relaxation (SOR) scheme. We refer to the original papers \citet{mina2019scalar} and \citet{2002AA...385..337T} for further details on the implementation of numerical schemes.

\section{Results}\label{sec:results}

\begin{figure*}
\centering
\includegraphics[width=\linewidth]{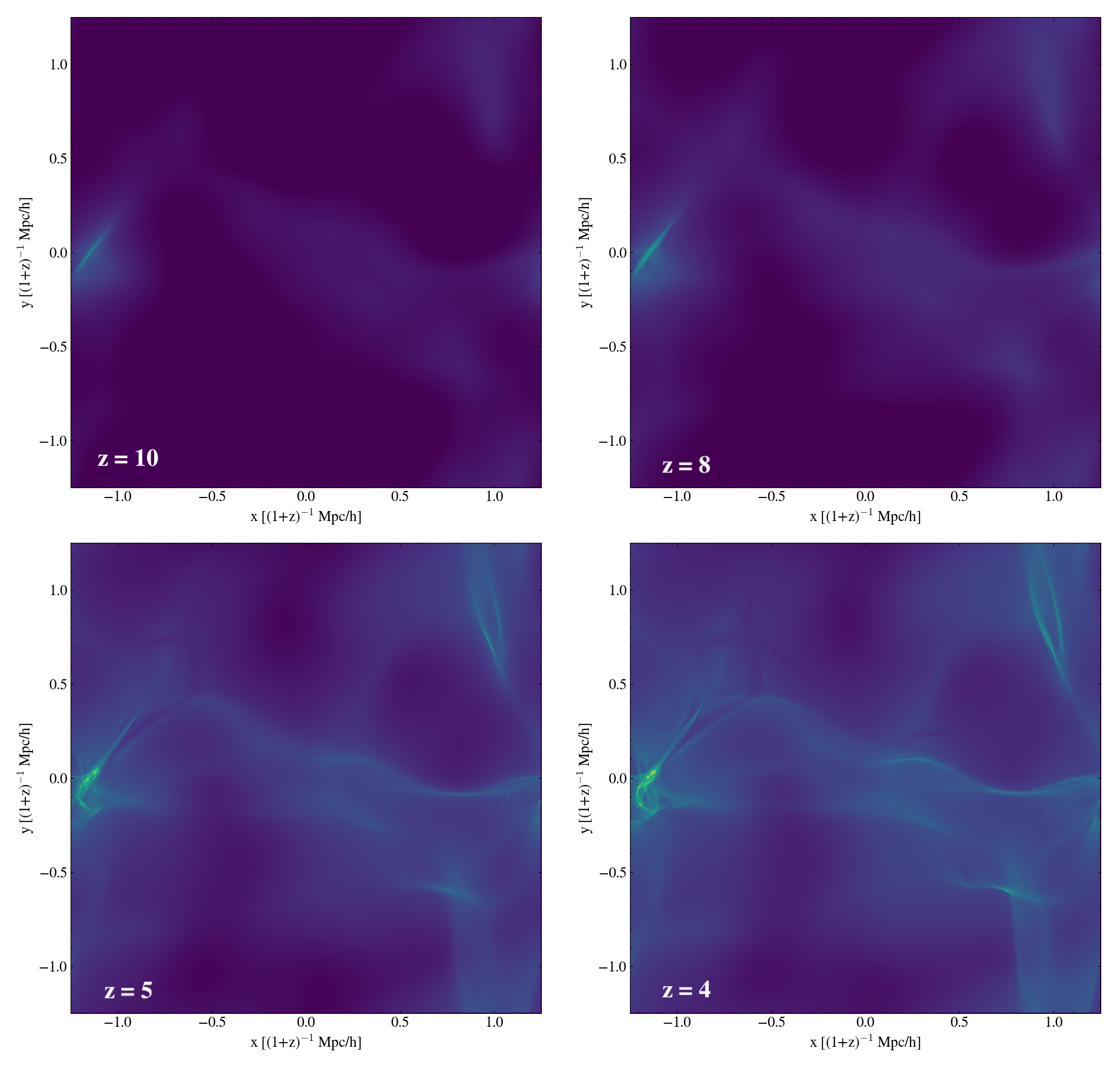}
\caption{
Projection along the $z$ axes of the dark matter density field, normalised by the critical density of the Universe. The box is $2.5~\text{Mpc}/h$ in comoving units and it represents the entire simulation box.
}\label{Fig:Evolution}
\end{figure*}

\begin{figure}
\centering
\includegraphics[width=\columnwidth]{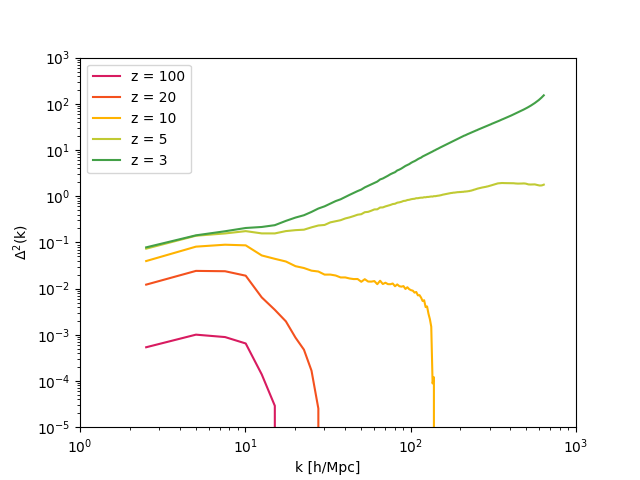}
\caption{
Evolution of the dimensionless power spectrum $\Delta^2(k)$ with redshift.
}\label{Fig:Pofk}
\end{figure}

\begin{figure*}
\centering
\includegraphics[width=\linewidth]{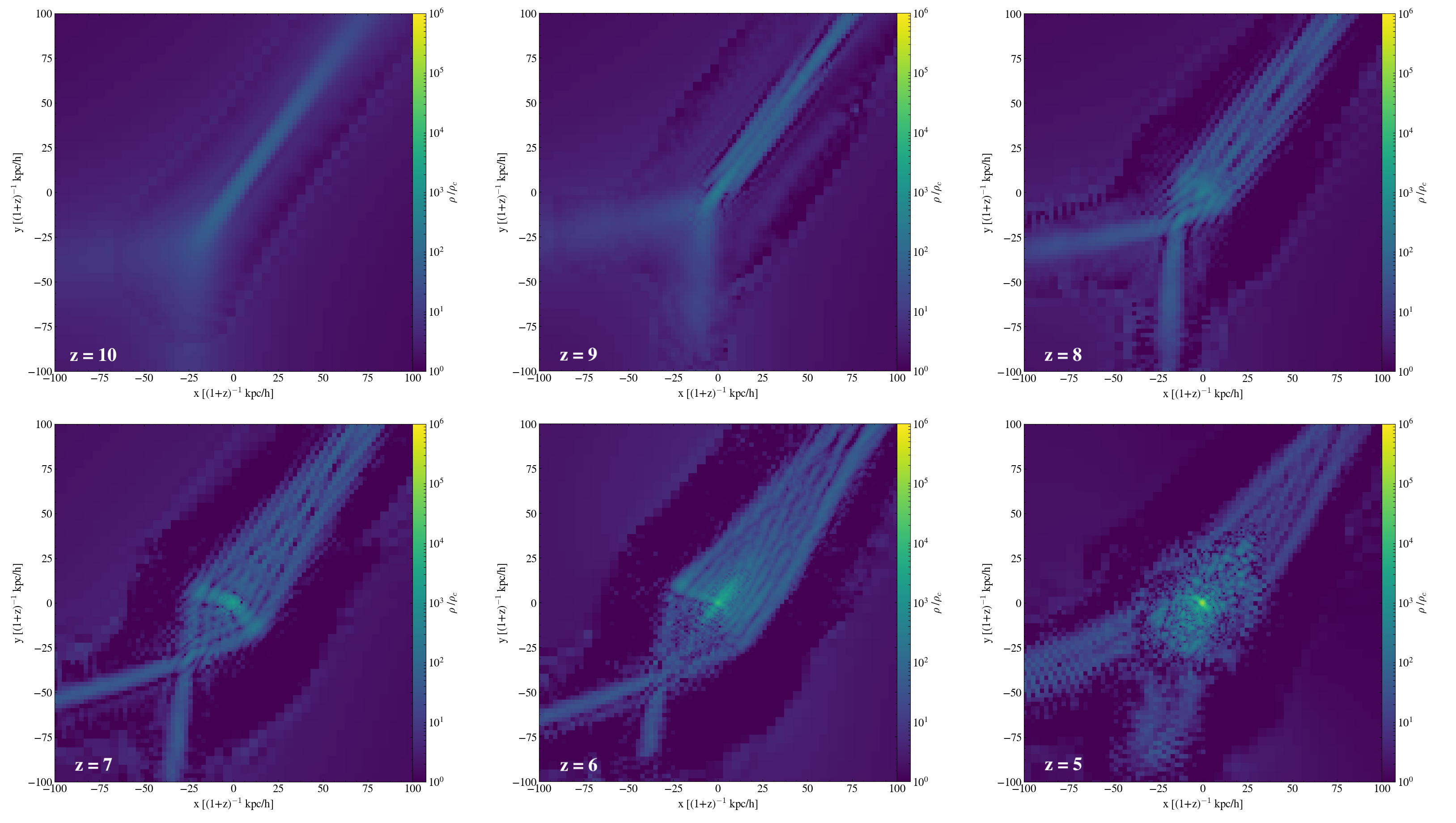}
\caption{
Slice along the $z$ axis, across one of the first structures that formed in our simulation box. Each panel represents a slice of $200~h^{-1}$kpc on a side, in comoving units. The density is normalised by the critical density of the Universe.
}\label{Fig:Formation}
\end{figure*}

\begin{figure*}
\centering
\includegraphics[width=\linewidth]{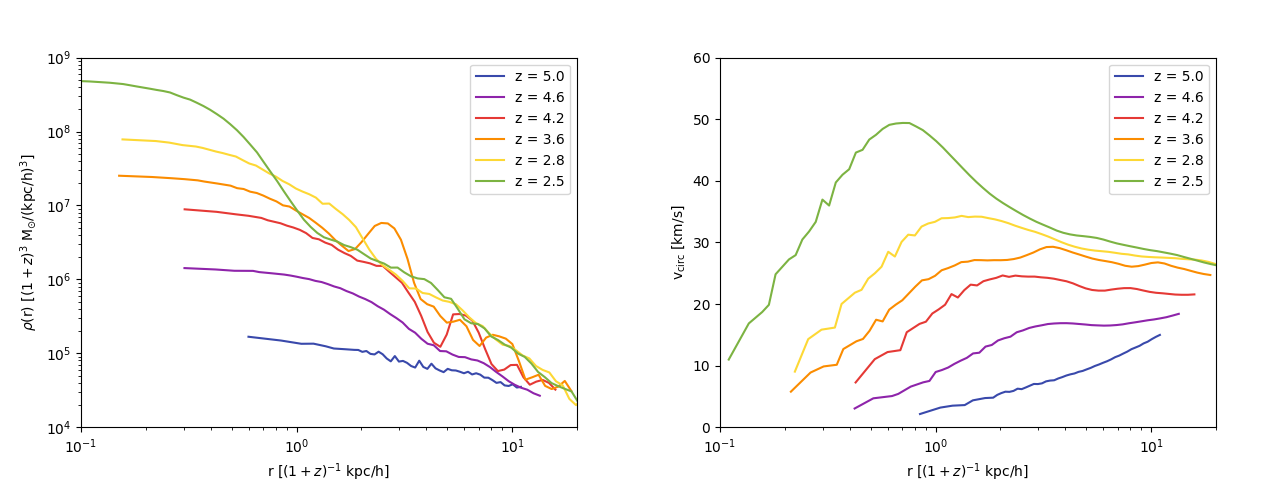}
\caption{
Comoving density (left panel) and circular velocity (right panel) profiles of \textsc{Halo 2} at redshifts $z=5.0$, $4.6$, $4.2$, $3.6$, $2.8$ and $2.5$.
}\label{Fig:Halo2_prof}
\end{figure*}

\begin{figure}
\centering
\includegraphics[width=\linewidth]{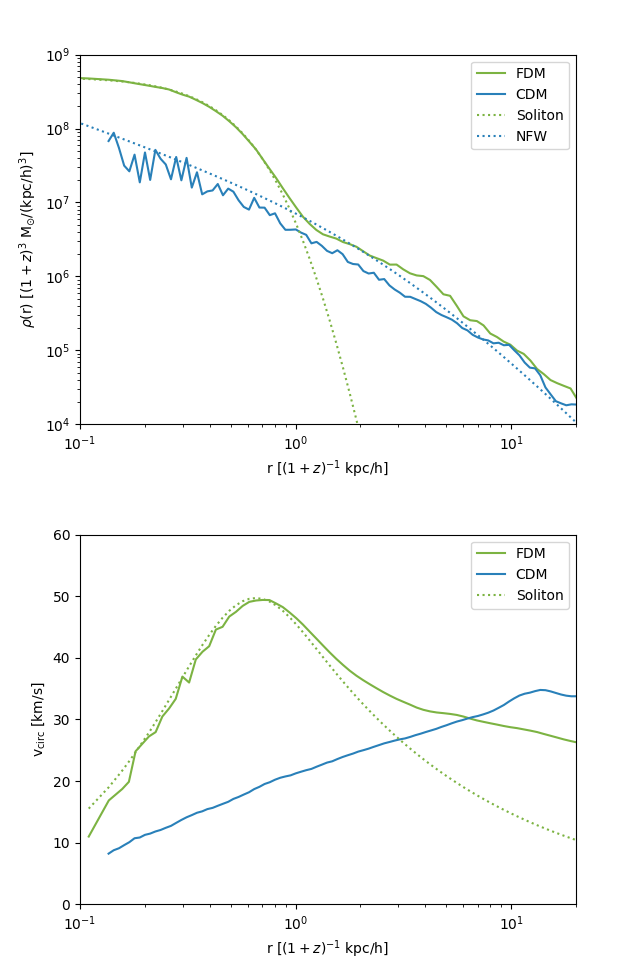}
\caption{
Density (upper panel) and circular velocity (bottom panel) profiles of \textsc{Halo 2} at redshift $z=2.5$. The solid lines show the FDM (green) and the CDM (blue) halos. While the dotted green line represents result of the fit of the solitonic core found in the FDM halo, the dotted blue line represents the result of the NFW fit of the CDM halo.
}\label{Fig:Halo2_prof_fin}
\end{figure}

\begin{figure*}
\centering
\includegraphics[width=\linewidth]{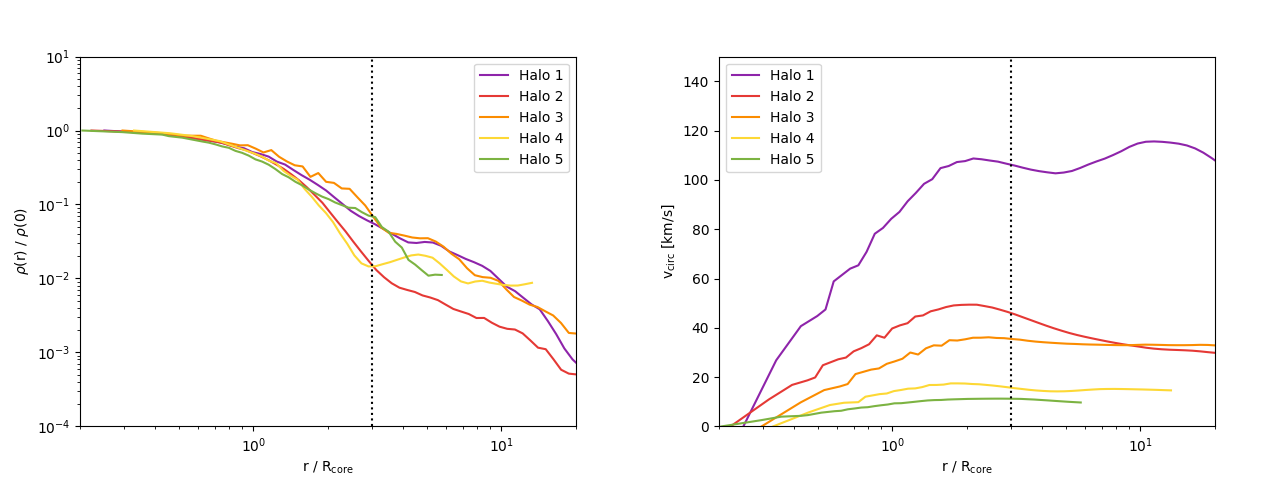}
\caption{
Dark matter density profiles (left panel) and rotation curves (right panel) of a representative sample of five FDM halos found at redshift $z=2.5$. The density if each individual halo is expressed in units of its central density. Similarly, the radial distance is normalised to the core radius of each individual halo. In both panels, the vertical dashed line corresponds to $r = 3 R_{\rm core}$, marking the point where the solitonic profile breaks.
}\label{Fig:Selection}
\end{figure*}

\begin{figure}
\centering
\includegraphics[width=\linewidth]{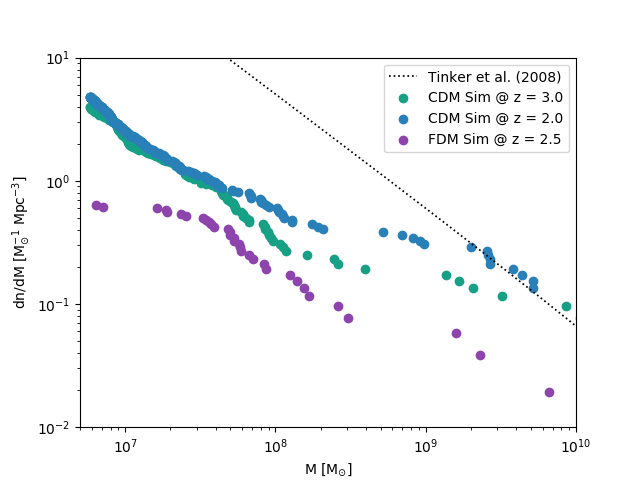}
\caption{
Halo Mass Function (HMF) computed for the FDM and CDM simulations. The black dotted line corresponds to the analytical estimate of the HMF in a $\Lambda$CDM cosmology, based on \citet{2008ApJ...688..709T}.
}\label{Fig:HMF}
\end{figure}

\begin{figure}
\centering
\includegraphics[width=\linewidth]{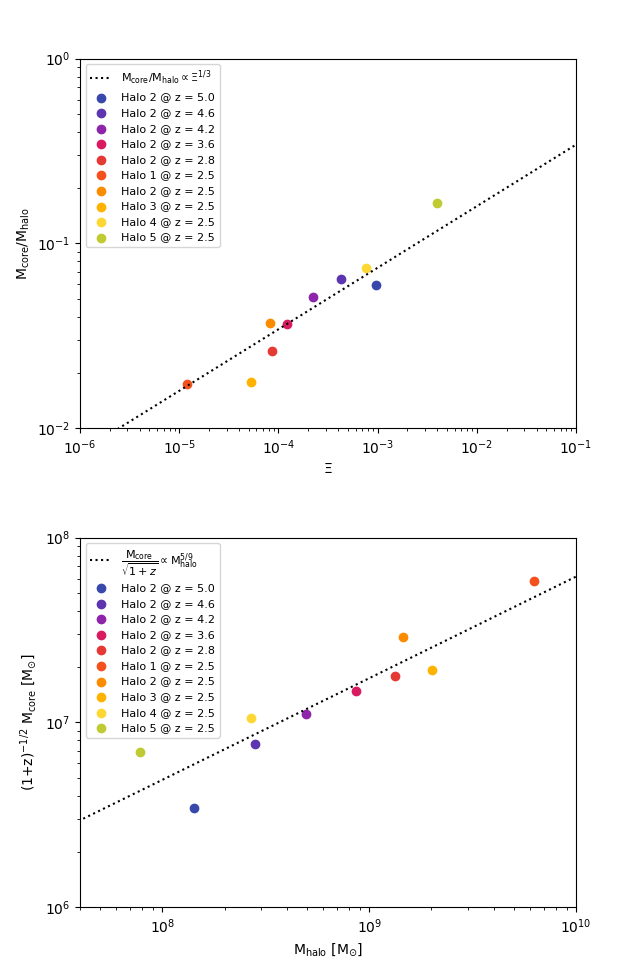}
\caption{
Core--halo mass scaling relation for a representative selection of five FDM halos.
}\label{Fig:sample-mcmh}
\end{figure}

\begin{figure}
\centering
\includegraphics[width=\linewidth]{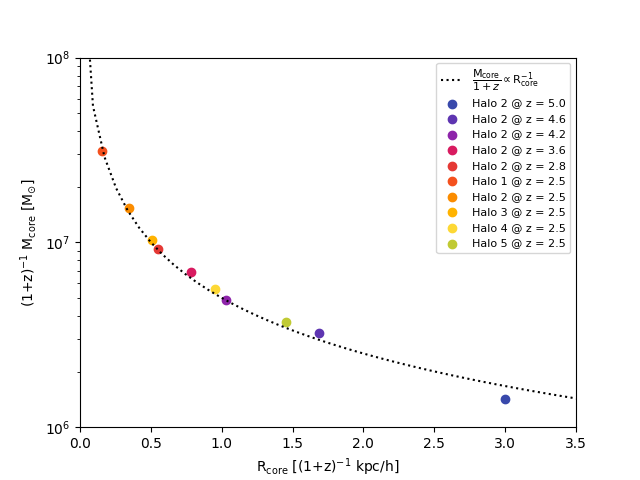}
\caption{
Core mass--radius scaling relation for a representative selection of five FDM halos.
}\label{Fig:sample-mcrc}
\end{figure}

\begin{figure}
\centering
\includegraphics[width=\linewidth]{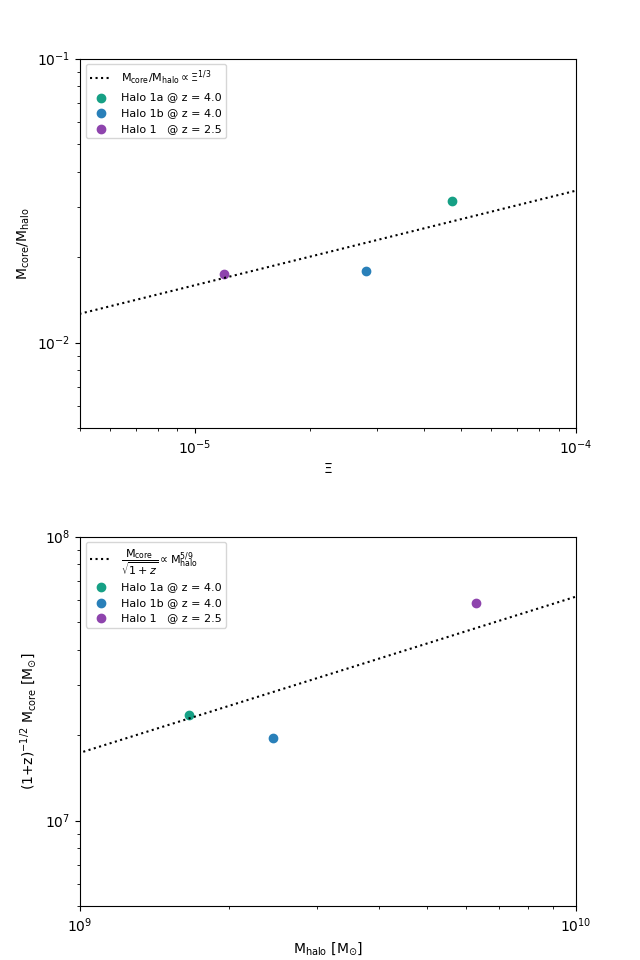}
\caption{
Core--halo mass scaling before and after the merging event of two FDM halos.
}\label{Fig:merger-mcmh}
\end{figure}

\begin{figure}
\centering
\includegraphics[width=\linewidth]{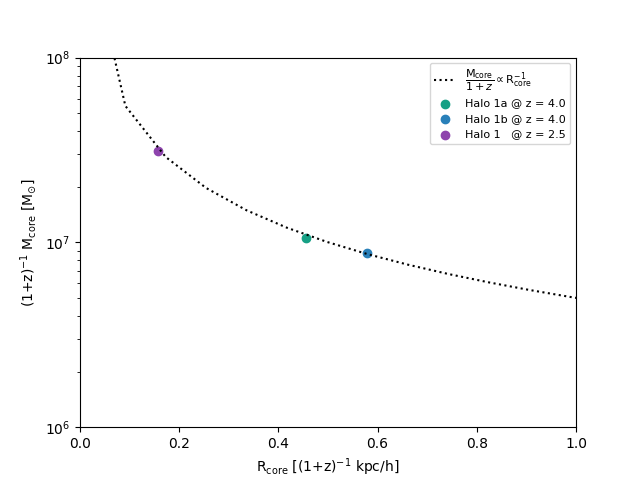}
\caption{
Core mass--radius scaling relation before and after the merging event of two FDM halos.
}\label{Fig:merger-mcrc}
\end{figure}

\begin{figure}
\centering
\includegraphics[width=\linewidth]{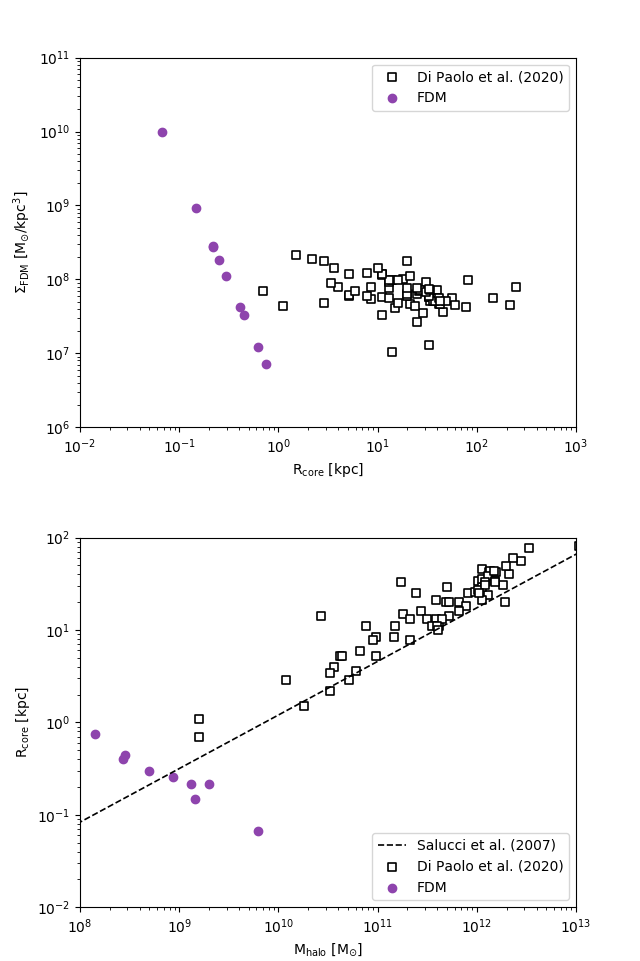}
\caption{
Upper panel: relation between the dark matter surface density and the core radius of dark matter halos. 
Bottom panel: relation between the core radius and the virial mass of dark matter halos. In both panels, the purple circles represent the simulated FDM halos, while the empty squares represent the properties of dark matter halos hosting observed LSB dwarf galaxies \citep{2020arXiv200503520D}. In addition, the black dashed lines corresponds to the result of the universal rotation curve (URC) method presented in \citet{10.1111/j.1365-2966.2007.11696.x}.
}\label{Fig:burkert}
\end{figure}

In this section we present the results of our simulation. The analysis is partially made with a modified version of the YT python package \citep{2011ApJS..192....9T}, and the halo catalogs are computed by using the Amiga Halo Finder \citep{Knollmann_2009}. A virial overdensity $\Delta_{\rm vir} = 200$ is used when locating halos.

\subsection{General evolution of a FDM Universe}

In Fig.~\ref{Fig:Evolution}, we project along the line-of-sight the FDM density field, normalised by the critical density of the Universe. 
The two-point correlation function of matter density perturbations is described by the matter power spectrum. 
By considering the Fourier transform of the density contrast:
\begin{align}
\delta(\mathbf{k}) = \int d^3 x ~ \delta(\mathbf{x}) \exp (-i\mathbf{k} \cdot \mathbf{x} ) ~ ,
\end{align}
the power spectrum $P(k)$ can be defined in terms of the autocorrelation function:
\begin{align}
\braket{ \delta(\mathbf{x}) \delta(\mathbf{x+r}) } &= \int_{0}^{\infty} \frac{dk}{k} \frac{k^3 P(k)}{2\pi^2}\frac{\sin(kr)}{kr} ~ ,
\end{align}
or, alternatively, in its dimensionless form:
\begin{align}
\Delta^2(k) = \frac{k^3 P(k)}{2\pi^2}~.
\end{align}
For $k \ll 1$ ${h~\rm Mpc}^{-1}$ the power spectrum probes modes in the density field that are still in the linear regime, thus describing the large-scale structure of the Universe. On the other hand, at $k \gg 1$ ${h~\rm Mpc}^{-1}$, the power spectrum encodes information about the evolution of the Universe at galactic and subgalactic scales, evolving in the non-linear regime. 

Due to the quantum nature of the scalar field, the FDM model predicts a sharper cutoff on non-linear scales than the $\Lambda$CDM model. The quantum pressure generated in high density regions and on scales of the de Broglie wavelength suppresses the formation of structures above the corresponding Jeans wavenumber, which can be estimated by linear theory as:
\begin{align}
k_{\rm J} &= \frac{66.5} {(1+z)^{1/4}} \left ( \frac{\Omega_{\rm FDM} h^2}{0.12} \right )^{1/4} \left ( \frac{m_{\rm B}}{10^{-22}~{\rm eV}} \right )^{1/2} ~ {\rm Mpc}^{-1} ~ ,
\end{align}
where $\Omega_{\rm FDM}$ corresponds to the present day dark matter relic abundance. 
Thus, the redshift dependent comoving Jeans wavelength $\lambda_{\rm J} = 2\pi/k_{\rm J}$ is given by:
\begin{align}
\lambda_{\rm J} &= \frac{ (1+z)^{1/4}}{10.6} \left ( \frac{\Omega_{\rm FDM} h^2}{0.12} \right )^{-1/4} \left ( \frac{m_{\rm B}}{10^{-22}~{\rm eV}} \right )^{-1/2} ~ {\rm Mpc} ~ ,
\end{align}
As opposed to the CDM case, the presence of a Jeans wavelength in the FDM model is directly connected to the quantum pressure.
As a consequence, below scales comparable with the Jeans wavelength, structure formation is heavily suppressed. 
Thus, the corresponding Jeans mass can be defined as follows:
\begin{align}\label{eq:massj}
M_{\rm J} &= \frac{4}{3}\pi \rho_{\rm B} \left ( \lambda_{\rm J}/2 \right )^3 =  \nonumber \\
&= 1.47\times 10^7 (1+z)^{3/4} \left ( \frac{\Omega_{\rm FDM} h^2}{0.12} \right )^{1/4} \left ( \frac{m_{\rm B}}{10^{-22}~{\rm eV}} \right )^{-3/2} ~ M_{\odot} ~.
\end{align}
Assuming that the solitonic core extends to the virial radius of the dark halo, Eq.\eqref{eq:massj} provides an lower bound on the mass of FDM halos.

In Fig.~\ref{Fig:Pofk}, we plot the evolution of the dimensionless power spectrum $\Delta^2(k)$ with redshift. In the linear regime (i.e. early times and large scales) the power-spectrum evolve as predicted by linear theory:
\begin{align}
\Delta^2_{\rm lin} \propto (1+z)^{-2} ~ ,
\end{align}
as for the CDM case.

\subsection{A dark halo growing in a FDM Universe}\label{halo-evo}

In our simulation, the first structures start forming around redshift $z\sim 10$. 
In Fig.~\ref{Fig:Formation}, we represent the dark matter field, normalised by the critical density of the Universe, and we show how it evolves as it forms one of the first structures. 
Starting from a small clump in form of a filament, dark matter is accreted more and more towards the deepest point of the gravitational potential well. As the system evolves towards a coherent state, the macroscopic wave-function develops the first interference fringes, revealing for the first time the quantum nature of the dark matter fluid. 
In the region around the minima of the gravitational potential, density grows over time forming a small gravitationally-bound structure: the first FDM halo. At the same time, on scales comparable with the de Broglie wavelength of the scalar field, the quantum pressure builds up as the scalar field collapses further, pushing energy from the highest density region of the dark halo to its outskirts.
A coherent and stable configuration develops in the innermost region of the collapsed object, forming a soliton with nearly constant density at the center. 
Outside the soliton radius, the field acquires its typical granular structure, which significantly differs from the classical CDM halo. These small coarse-grained clumps are originated by the superposition of multiple plane-waves, resulting from the flow of energy promoted by the quantum pressure.

For illustrative purposes, we follow the evolution a single FDM halo during the expansion of the Universe. 
In this regard, we select an isolated halo, named \textsc{Halo 2}, among the most massive ones formed in our simulation. As shown in Fig.~\ref{Fig:Halo2_prof}, we track its density profile as the scalar field collapses under the effect of gravity.
Due to the quantum nature of the scalar field, the innermost region of the dark halo exhibits a solitonic core with an almost constant density at all redshifts, which characterises the soliton solution of the non-linear Schr\"odinger equation. 
The solitonic core extends on scales corresponding to the coherence length of the scalar field, or its de Broglie wavelength. On larger scales, where a high level of coherence is not achieved, the density profile quickly drops as a power-law. 

In Fig.~\ref{Fig:Halo2_prof_fin}, we only consider \textsc{Halo 2} at redshift $z=2.5$. In the upper panel, we fit the innermost region of its density profile with the soliton profile:
\begin{align}\label{eq:soliton}
\rho(r) &= 1.9 \times 10^9 h a^{-1} \frac{\left ( \dfrac{m_{\rm B}}{10^{-23}~{\rm eV}} \right )^{-2} \left (\dfrac{R_{\rm core}}{{\rm kpc/h}} \right )^{-4} }{ \left [1+9.1\times 10^{-2} \left (\dfrac{r}{R_{\rm core}} \right )^{2} \right ]^8} ~ \frac{M_{\odot}}{({\rm kpc/h})^3} ~,
\end{align}
leaving the core radius as the only free parameter. This formula was first suggested by \citet{2014NatPh..10..496S} and it defines the core radius as the point where the density drops by half of its central value. In this case, the solitonic core is well fitted by Eq.~\eqref{eq:soliton} with a core radius of $R_{\rm core} = 0.35$ $h^{-1}$ kpc, in comoving units.
At the same time, we fit the density profile at larger radial distances with the NFW profile, given by Eq.~\eqref{eq:nfw}. In this case, the free parameters of the fit are the central density $\rho_0$ and the concentration of the halo, defined as $c_{\rm NFW} = R_{\rm vir}/R_{\rm s}$. 
Assuming the same virial radius found for the FDM halo, the best fit yields an NFW profile characterised by a comoving central density of $\rho_0 = 4.26\times 10^6$ $M_{\odot}$ $(h^{-1}$ kpc$)^{-3}$ and a concentration parameter of $c_{\rm NFW} = 10.08$. 
On the bottom panel of Fig.~\ref{Fig:Halo2_prof_fin}, instead, we plot the circular velocity of the FDM halo, which is computed according to the formula:
\begin{align}\label{eq:circ-vel}
v_{\rm circ}(r) = \sqrt{\frac{G M_{\rm enc}}{r}} ~ ,
\end{align} 
where $G$ denotes the Newton's gravitational constant and $M_{\rm enc}$ corresponds to the mass enclosed within the radius $r$. 

By redshift $z=2.5$, the $\Lambda$CDM Universe simulated on a side has produced a huge amount of structures and substructures, making a direct matching between FDM and CDM halos an impossible task. For this reason, we select a dark halo in the simulated $\Lambda$CDM Universe with virial properties similar to \textsc{Halo 2} and, in Fig.~\ref{Fig:Halo2_prof_fin}, we plot its density profile and rotation curve. 
The circular velocity profile of the CDM halo is characterised by a single peak located at $R_{\rm max} = (\alpha / c_{\rm NFW}) R_{\rm halo}$, with $\alpha \sim 2.16$.
Instead, in the case of the FDM halo, the circular velocity peak is located at much smaller radii than the CDM halo, due to the presence of a small compact and prominent solitonic core at the center of the dark halo. 
In other cases the circular velocity profiles of FDM halos exhibit two peaks, one connected to the solitonic core at the center and one connected to the NFW-like outer region contribution, as discussed later in Section~\ref{low-z}.
In some cases, the presence of such a solitonic core can be disfavoured if circular velocity profiles are compared with real observed dwarfs \citep{2018PhRvD..98h3027B,2019PhRvD..99j3020B}.

Finally, at redshift $z=2.5$, \textsc{Halo 2} has a virial mass and a comoving virial radius of $M_{\rm vir} =1.56 \times 10^9 ~{\rm M}_{\odot}$ and $R_{\rm vir}=24.21 ~ h^{-1} {\rm kpc}$, respectively.

\subsection{Non-linear regime at low redshift}\label{low-z}

At redshift $z=2.5$, we select a sample of five representative halos found in our simulation box, spanning over three dex in virial mass.  
The density profiles of the five halos are plotted in the left panel of Fig.~\ref{Fig:Selection}, while in the right panel we plot their rotation curves computed according to Eq.~\eqref{eq:circ-vel}. By using the soliton profile, Eq.~\eqref{eq:soliton}, we fit the innermost region of the density profiles, estimating in this way the core radius of each halo. Apart from some scatter beyond the soliton radius due the complex dynamics, the outer part of each FDM halo decays in a similar way to the classical CDM halo, following an NFW profile. The transition between the soliton profile and the NFW outskirt is found to universally occur at a radius of $r\sim3R_{\rm core}$.
Typically, high-mass FDM halos are characterised by more prominent solitonic cores, with higher core masses and smaller core radii than low-mass FDM halos. This reflects the core mass--radius and core--halo mass scaling relations, which are investigated in details in Section~\ref{sec:scaling}.

By computing the moment of intertia tensor, we can estimate the shapes of the FDM halos, which can have important observational consequences. 
Contrary to the results obtained in \citet{Mocz_2020} for the case of BECDM, where dark halos were found to be more triaxial than typical CDM halos, we find that FDM halos follow the characteristic triaxial configuration of CDM halos, with axis ratios of $b/a \sim c/a \sim 0.55-0.85$.

In addition, for both the FDM and the CDM simulations, we compute the halo mass function (HMF) by counting the number of structures falling in a given mass bin. 
Due to the small size of the box, we can only probe mass scales up to $M\lesssim 10^{10}~M_{\odot}$, corresponding to typical dwarf masses.
In Fig.~\ref{Fig:HMF}, we compare the two HMFs with the $\Lambda$CDM estimate provided by \citet{2008ApJ...688..709T}.
First, we note that the smallest halo in the simulated FDM Universe at redshift $z=2.5$ have a virial mass of $7.14 \times 10^{6} ~M_{\odot}$, which is consistent with the minimum mass estimate provided by Eq.~\eqref{eq:massj}.
Then, we note that the HMF of the simulated FDM Universe is suppressed at all mass scales. 
In the CDM run, the HMF also shows signs of suppression at masses $M_{\rm halo} \lesssim 10^{9}~M_{\odot}$, but this is an artificial effect due to the fact that we employ the FDM initial conditions also for the CDM run (the initial linear power-spectrum is suppressed at small scales). However, below $M_{\rm halo} \lesssim 10^{8}~M_{\odot}$ the HMF computed for the CDM run exhibits the same trend as expected for the $\Lambda$CDM model, growing as: 
\begin{align}
\frac{dn}{dM} \propto M^{-2} ~ .
\end{align}
The stellar mass function (SMF) measured for observed galaxies is also suppressed at typical mass scales of dwarf galaxies. In this case, the observed suppression is attributed to various feedback mechanisms dominating in low-mass and dark matter dominated systems, resulting in strong star formation inefficiencies. 
Within the $\Lambda$CDM model, such a big difference between the HMF and the SMF is at the origin the missing satellite and too-big-to-fail \citep{2011MNRAS.415L..40B} problems. 
In the case of a FDM Universe, while the suppression of the HMF above $M_{\rm halo} \sim 10^8~M_{\odot}$ can be attributed to the lack of statistics, the suppression found in the low-mass end is a direct consequence of the quantum nature of the scalar field and it clearly shows the capability of this model to work in the right direction to solve the aforementioned small-scale problems.

Furthermore, the existence of a circular velocity peak due to a small compact core, shown in Fig.~\ref{Fig:Halo2_prof_fin} and Fig.~\ref{Fig:Selection}, can have important consequences for the FDM model \citep{2018PhRvD..98h3027B}. 
If baryonic physics was included in the treatment, small compact solitonic cores could, in principle, enhance gravitational cooling and accretion of gas towards the center of dark matter halos in systems with virial masses above a critical mass of $M_{\rm halo} \gtrsim 10^{8} ~ M_{\odot}$. As suggested by \citet{Schive_2014}, formation of ultra-dense gas in the center of the dark halo could promote major starburst and early forming quasars \citep{f2293ac541c24ba690492c90645c960e}.
On the other hand, due to the bursty star formation observed in dwarf galaxies, stellar feedback often prevents gas accretion, leading to a complex interplay between gravitational cooling and heating processes resulting from various feedback mechanisms. 
As a matter of facts, baryonic processes are known to be of extreme importance even in dark matter dominated systems, such as dwarf galaxies, and therefore we cannot draw any conclusion based only on the results obtained from our simulated FDM Universe.

We reserve for a future work a proper numerical investigation including baryonic physics, as well as relevant astrophysical processes such as stellar feedback.

\subsection{Scaling relations}\label{sec:scaling}

Since the first numerical studies about the formation and the evolution of FDM halos, it was shown that general properties of FDM halos are tightly linked by a series of interesting scaling relations, resulting from intrinsic scaling symmetries of the Schr\"odinger-Poisson system.
Assuming no net angular momentum, a FDM halo is primarily characterised by a single dimensionless parameter:
\begin{align}\label{eq:Xidef}
\Xi = \frac{ \left | E_{\rm halo} \right | / M_{\rm halo}^3}{ m_{\rm B}G } ~ ,
\end{align}
which is a scale-free invariant of the Schr\"odinger--Poisson system \citep{2017MNRAS.471.4559M,2018PhRvD..98h3027B}. In Eq.~\eqref{eq:Xidef}, $M_{\rm halo}$ corresponds to the viral mass of the dark halo, while $ \left | E_{\rm halo} \right |$ is its total energy, which can be approximated as:
\begin{align}
\left | E_{\rm halo} \right | \sim \frac{G M_{\rm halo}^2}{R_{\rm halo}} ~ .
\end{align}
As in other numerical studies, we find a fundamental scaling relation between the core mass $M_{\rm core}$ and $\Xi$ in the form:
\begin{align}\label{eq:scaling1}
M_{\rm core}/M_{\rm halo} = \alpha ~ \Xi^{\beta} ~ .
\end{align}
This means that, given its initial mass and energy, each FDM halo can be uniquely described by the $\Xi$ parameter. 
In order to quantify the scaling of the core mass with the scale-free invariant $\Xi$, we use Eq.~\eqref{eq:scaling1} to fit a data sample containing the selection of five halos described in Section~\ref{low-z}, together with different temporal realisations of \textsc{Halo 2} presented in Section~\ref{halo-evo}. 
Including the $1-\sigma$ error on the parameters, the fit yields to $\alpha = 1.21 \pm 0.162$ and $\beta = 0.39 \pm 0.043$, which is consistent with the results of \citet{2017MNRAS.471.4559M} \citep[see also][]{2016PhRvD..94d3513S,2017PhRvD..95d3519D}. 
Thus, fixing the parameter $\beta = 1/3$, the core--halo mass relation is given by:
\begin{align}\label{eq:scaling1b}
M_{\rm core}/M_{\rm halo} = 0.73 ~ \Xi^{1/3} ~ ,
\end{align}
and it is shown in the upper panel of Fig.~\ref{Fig:sample-mcmh}.
By further approximating the total energy of the halo as $\left | E_{\rm halo} \right | \propto M_{\rm halo}^{5/3}$, Eq.~\eqref{eq:scaling1b} would imply a scaling of $M_{\rm core} \propto M_{\rm halo}^{5/9}$ between the core and the halo masses. To check the validity of this approximation, we also fit the same data sample with:
\begin{align}\label{eq:scaling2}
M_{\rm core} = 10^{\alpha} \left (1+z \right )^{1/2}\left(\frac{M_{\rm halo}}{M_{\odot}}\right)^{\beta}M_{\odot}~,
\end{align}
where we explicitly take into account the redshift dependence suggested by \citet{Schive_2014}. As expected, the fit yields to $\alpha = 2.29 \pm 0.709$ and $\beta = 0.55 \pm 0.081$. 
In the lower panel of Fig.~\ref{Fig:sample-mcmh}, we show the alternative form of the core--halo mass relation provided by Eq.~\eqref{eq:scaling2}, and we note that the biggest scatter among the data corresponds to \textsc{Halo 2} at redshift $z=5$ and to \textsc{Halo 5} at redshift $z=2.5$, which are characterised by small masses and might be not fully virialised yet.
Other studies have found a similar core--halo mass relation, but with different values for the $\beta$ exponent in Eq.~\eqref{eq:scaling2}. For example, in \citet{2018PhRvD..97h3519M} it was found a value of $\beta = 1/9$, while in \citet{Schive_2014} it was found a value of $\beta = 1/3$. 
The core--halo mass scaling has interesting implications. As mentioned by \citet{Schive_2014}, the uncertainty principle in quantum mechanics is a local relation, but the core--halo mass relation links a local property like the core mass to a global property such as the virial mass of the halo, implying that the uncertainty principle holds, in this case, non-locally.

At the same time, the core radius and its mass are tightly connected by the core mass--radius relation. With our definition of core radius, the solitonic core encloses roughly $25$\% of the total soliton mass. As a consequence, the core mass can be expressed as:
\begin{align}\label{eq:mass-soliton}
M_{\rm core} = 2.1 \times 10^{10} h a^{-1} \left ( \dfrac{m_{\rm B}}{10^{-23}~{\rm eV}} \right )^{-2} \left (\dfrac{r_{\rm core}}{{\rm kpc/h}} \right )^{-1} ~ M_{\odot} ~ .
\end{align} 
This equation can be obtained by integrating the soliton profile, given by Eq.~\eqref{eq:soliton}. In Fig.~\ref{Fig:sample-mcrc}, we show how the core mass computer for our sample of FDM halos scales accordingly to the core radius.

\subsection{Merging two FDM halos}

The merging process between FDM matter halos have not been studied before in a realistic cosmological environment. 
Initially, collisions between self-gravitating solitons were studied in \citet{PhysRevD.74.103002}. In this work, it was shown that if the total initial energy of a binary system is positive, the two solitons passes through each other, while in case of a negative total initial energy the two soliton merge into one virialised structure.
In \citet{Schive_2014},\citet{2017MNRAS.471.4559M} and \citet{2016PhRvD..94d3513S}, the merger between solitons have been reproduced from an idealised set of initial conditions and it was shown that the core mass--radius and the core--halo mass scaling relations are preserved by the merging process.
Furthermore, \citet{2016PhRvD..94d3513S} showed that the core mass resulting from a binary merger only depends on the mass ratio and on total initial mass and energy, while it is independent on the initial phase difference and angular momentum. In addition, when the progenitors have non-zero angular momentum the final core becomes a rotating ellipsoid, which otherwise would be spherical.

In our simulation, the two most massive FDM halos formed by redshift $z=4$, namely \textsc{Halo 1a} and \textsc{Halo 1b}, start to slowly approach each other. By redshift $z\sim3$ they collide head-on and they merge into a single big dark matter halo, namely \textsc{Halo 1}. 
The two progenitors of \textsc{Halo 1} have very similar virial properties: with masses of $M_{\rm 1a}=1.9 \times 10^9$ $M_{\odot}$ and $M_{\rm 1b}=2.9 \times 10^9$ $M_{\odot}$, respectively, the merger event would be classified as a major merger.
The final product, \textsc{Halo 1}, is a larger dark matter halo with virial mass and comoving virial radius of $M_{\rm 1}=6.5 \times 10^9$ $M_{\odot}$ and $R_{\rm 1}=39.2 ~h^{-1}$kpc, respectively. 

The core radius of the progenitors \textsc{Halo 1a} and \textsc{Halo 1b} are computed by fitting with Eq.~\eqref{eq:soliton} their soltonic cores. At redshift $z=4$, right before the merging event starts, we find $R_{\rm core,1a}=0.45 ~h^{-1}$kpc and $R_{\rm core,1b}=0.58 ~h^{-1}$kpc in comoving units, for \textsc{Halo 1a} and \textsc{Halo 1b} respectively. At redshift $z=2.5$, the final dark matter halo have a comoving core radius of $R_{\rm core,1}=0.16 ~h^{-1}$kpc. As shown in Fig.~\ref{Fig:merger-mcmh} and Fig.~\ref{Fig:merger-mcrc}, we also find that both the core mass--radius and the core--halo mass relations are preserved by the merging process.
With a mass ratio between the progenitors of approximately $\mu = M_{\rm 1b}/M_{\rm 1a} \sim 1.5$, the core mass of the most massive one is enhanced by the merger, in agreement with \citet{2016PhRvD..94d3513S}, where it was shown that disruption of the least massive core is achieved only for mass ratios above $\mu > 7/3$ and in this case the least massive core is dispersed in the NFW-like outskirt of the most massive halo. 

\subsection{Core surface density}
Dark matter cores have been mostly observed for low-mass galaxies, with stellar masses of $M_{*} \lesssim 10^{10}~M_{\odot}$ and small baryon fractions \citep{2020ApJS..247...31L,2015PNAS..11212249W,2020arXiv200503520D}. Density profiles of such low-mass dark halos are commonly described by means of an isothermal profile
\citep{1995ApJ...447L..25B}: 
\begin{align}\label{eq:pseudoiso}
\rho_{\rm{ISO}}(r) = \rho_0 \frac{r_0^3}{\left(r+r_0\right ) \left( r^2+r_0^2 \right)} ~ ,
\end{align}
which provides a better fit to the observed data. In Eq.~\eqref{eq:pseudoiso}, free parameters $\rho_0$ and $r_0$ correspond to the central density and the core radius, respectively.

Observations of dark matter cores have shown that the central density and the core radius are tightly related by the core surface density $\Sigma_{\rm DM} = \rho_0~r_{0}$ \citep[e.g.][]{2000ApJ...537L...9S,2015ApJ...808..158B,2015llg..book..323K,2009MNRAS.397.1169D}. This relation can provide strong constraints on any core formation mechanism and, in \citet{2020arXiv200611111B}, it was recently argued that this represents a challenge to the FDM model to explain the observed cores. 
In Fig.~\ref{Fig:burkert}, we compare our simulated FDM halos with the properties of dark matter halos hosting low surface brightness (LSB) dwarf galaxies \citep{2020arXiv200503520D}. We caution that a proper comparison would require a connection of the halo mass to the observed stellar mass, and that we only aim at illustrate the general trend found in our simulation for FDM halos.
As shown in the upper panel of Fig.~\ref{Fig:burkert}, observed dwarf galaxies are highly consistent with a constant core surface density, with an average of $\Sigma_{\rm DM} = 75~M_{\odot}$pc$^{-2}$.
However, in the case of FDM, the trend is drastically different. Indeed,  Eq.~\eqref{eq:soliton} implies a scaling between the central core surface density and the core radius of $\Sigma_{\rm DM} \propto R_{\rm core}^{-3}$. 
In the bottom panel of Fig.~\ref{Fig:burkert}, we show the dependence of the core radius on the virial mass of the halo. Whereas the observed data exhibits a positive scaling between the $R_{\rm core}$ and $M_{\rm halo}$, our simulated FDM halos follow the relation $R_{\rm core} \propto M_{\rm halo}^{-5/9}$.

\section{Conclusions}\label{sec:conc}
Within the $\Lambda$CDM framework, the process of structure formation has been extensively studied in the past decades. 
Numerical simulations have shown that the collisionless nature of CDM allows dark matter to form structures at all probed masses, and that dark matter clusters in gravitationally-bound halos following the universal NFW density profile, Eq.~\eqref{eq:nfw}. 
However, observations of rotation curves in small, compact and dark matter dominated systems, such as dwarf galaxies, indicate that the innermost region of dark matter halos deviates from the NFW density profile, rather forming a small central core with nearly constant density. 
In addition, cosmological simulations have pointed out a strong mismatch between the number of observed low-mass subhalos and the number of simulated structures.
Thus, in the recent past, a variety of alternative dark matter model has been suggested to better describe the properties of observed structures. A promising alternative to the standard CDM is provided by models where the dynamics of dark matter is described by means of an ultra-light scalar field, such as in the FDM model. High resolution non-linear simulations are therefore required to test these models and to map out their unique signatures.

For this work, we simulate a $2.5~ h^{-1}$Mpc box representing a small portion of the Universe where the whole dark matter budget is in form of FDM, and it is described by a light complex scalar field with a mass of $m_{\rm B} = 2.5\times 10^{-22}~{\rm eV}$. 
While the dynamics of the scalar field is very similar to the CDM case on large scales, it significantly differs from standard CDM at small scales, due to the fact that the quantum nature of the scalar field is manifested on astronomically relevant scales.

We study the formation and evolution of FDM halos and show how the dynamics of a scalar field with such a small mass have many observational consequences, which can be used to probe the true nature of dark matter in a cosmological context.
Among different structures formed in our simulation, we select a representative sample of five FDM halos at redshift $z=2.5$, within a mass range $10^8 < M_{\rm halo} < 10^{10}~M_{\odot}$. 
Due to the very high resolution employed, we are able to resolve the innermost structure of dark halos and we show that FDM particles condense in the very center of each structure, forming a coherent solitonic core.
The core radius depends upon the mass of the corresponding soliton, following the well known core mass--radius relation. 
The central density profile is nearly constant on scales of the coherence length of the scalar field and. While the solitonic core is well approximated by the soliton profile given by Eq.~\eqref{eq:soliton}, the NFW-like outer region decays with a log-slope of $\gamma \sim -3$, similarly to CDM halos.

We also select one FDM halo among the ones in the sample, and we study the formation of the central soliton. We show how the solitonic core evolves as the halo collapses under the effect of gravity and we analyse its impact on the circular velocity profile of the dark halo. A central compact core leads to a prominent circular velocity peak in the rotation curve, at much smaller radii from the center than in the CDM case.

Furthermore, we characterise each dark halo in terms of the scale-free invariant $\Xi$, and we find that the simulated FDM halos follow the core--halo mass relation $M_{\rm core}/M_{\rm halo} \propto \Xi^{\gamma}$, characterised by a scaling exponent of $\gamma = 1/3$.
The form of this scaling relation has been a critical investigation point for this class of alternative dark matter model, as different numerical studies found different scaling exponents for the core--halo mass relation. Our results are well in line with \citet{2017MNRAS.471.4559M}, where the same core--halo mass relation was found for the case of BECDM.

In addition, mergers between FDM halo had never been studied in a realistic cosmological scenario. In our simulated FDM Universe, two among the most massive dark halos undergo a major merger between redshifts $z \sim 4$ and $z \sim 3$, and we find that the merging process preserve both the core mass--radius and core--halo mass scaling relations. 

Moreover, we compare the core surface density $\Sigma_{\rm DM}$ of the simulated FDM halos with that of real dark halos hosting observed LSB dwarf galaxies. For the FDM case, we find that the trend in both the $\Sigma_{DM}-R_{\rm core}$ and the $R_{\rm core}-M_{\rm halo}$ relations are in tension with the observed properties of real dark halos. As argued in \citet{2020arXiv200611111B}, the negative $R_{\rm core}-M_{\rm halo}$ scaling found for the simulated FDM halos represents a challenge to the FDM scenario as the sole explanation for the observed dark matter cores.
To further investigate the $\Sigma_{\rm DM}-R_{\rm core}$ and the $R_{\rm core}-M_{\rm halo}$ relations, larger numerical simulations with the FDM model are required to analyse larger samples of FDM halos and to cover a wider range of virial masses. In addition, baryonic physics can potentially change the predictions of FDM only simulations in a non-trivial way. 
Thus, if the arguments provided by \citet{2020arXiv200611111B} holds, even if FDM is capable of forming large solitonic cores in the center of dark halos, our results, together with previous results in the literature, suggest that the origin of the observed dark matter cores in low-mass astrophysical systems might have to be searched for somewhere else.

One potential shortcoming of this work is that the refinement strategy adopted in our simulation employs a refinement criterion that is only based on a fixed density threshold, at all refinement levels. As a future improvement, we consider to further develop \codename by improving the refinement strategy, including a criterion based on the Jeans length of collapsing objects. Such a criterion would be similar to the one adopted here, but with the difference of having a density threshold which vary as the Universe expands. Since the Jeans scale of FDM halos is redshift dependent, a refinement criterion based on a constant density threshold does not guarantee enough resolution to resolve late-forming halos, which typically have smaller masses than early-forming FDM halos. 
With a refinement strategy based on the Jeans length of collapsing object, the resolution required by these late-forming FDM halos would be automatically achieved at any redshift and any mass scale. A refinement criterion based on the fluid velocity, as advocated by \citep{Schive_2014} to ensure that the field is well behaved in low-density regions, would also be an interesting investigation point, to verify the robustness of the results obtained in this kind of simulations. This is especially important considering that results from different groups, while reaching similar conclusions, have shown to slightly differ in their quantitative predictions.

In addition, we plan to investigate further the dynamics of FDM, by including in our description the baryonic content of the Universe. Modelling the gas physics, together with the astrophysical component of structure formation, is crucial to make clear predictions out of numerical simulations of structure formation. The \codename code includes the hydrodynamics solver required for the gas physics and a sub-grid model for stellar feedback, as they were already implemented in the \texttt{RAMSES} code. 

\hfill\newline

{\it As we were just about to submit this manuscript on the arXiv, \citet{2020arXiv200701316N} appeared, reporting new results from cosmological simulations with the FDM model. We reserve for a future version of this manuscript a direct comparison with their work.}

\begin{acknowledgements}
We thank the Research Council of Norway for their support. Computations were performed on resources provided by UNINETT Sigma2 -- the National Infrastructure for High Performance Computing and Data Storage in Norway. 
\end{acknowledgements}

\bibliography{references.bib}

\begin{thebibliography}{72}
\expandafter\ifx\csname natexlab\endcsname\relax\def\natexlab#1{#1}\fi

\bibitem[{Aarseth(2003)}]{aarseth_2003}
Aarseth, S.~J. 2003, Gravitational N-Body Simulations: Tools and Algorithms,
  Cambridge Monographs on Mathematical Physics (Cambridge University Press)

\bibitem[{{Aarseth} \& {Hoyle}(1964)}]{1964ApNr....9..313A}
{Aarseth}, S.~J. \& {Hoyle}, F. 1964, Astrophysica Norvegica, 9, 313

\bibitem[{{Bar} {et~al.}(2018){Bar}, {Blas}, {Blum}, \&
  {Sibiryakov}}]{2018PhRvD..98h3027B}
{Bar}, N., {Blas}, D., {Blum}, K., \& {Sibiryakov}, S. 2018, PRD, 98, 083027

\bibitem[{{Bar} {et~al.}(2019{\natexlab{a}}){Bar}, {Blum}, {Eby}, \&
  {Sato}}]{2019PhRvD..99j3020B}
{Bar}, N., {Blum}, K., {Eby}, J., \& {Sato}, R. 2019{\natexlab{a}}, PRD, 99,
  103020

\bibitem[{{Bar} {et~al.}(2019{\natexlab{b}}){Bar}, {Blum}, {Lacroix}, \&
  {Panci}}]{2019JCAP...07..045B}
{Bar}, N., {Blum}, K., {Lacroix}, T., \& {Panci}, P. 2019{\natexlab{b}}, \jcap,
  2019, 045

\bibitem[{Bernal \& Guzm\'an(2006)}]{PhysRevD.74.103002}
Bernal, A. \& Guzm\'an, F.~S. 2006, Phys. Rev. D, 74, 103002

\bibitem[{{Boylan-Kolchin} {et~al.}(2011){Boylan-Kolchin}, {Bullock}, \&
  {Kaplinghat}}]{2011MNRAS.415L..40B}
{Boylan-Kolchin}, M., {Bullock}, J.~S., \& {Kaplinghat}, M. 2011, \mnras, 415,
  L40

\bibitem[{Boylan-Kolchin {et~al.}(2012)Boylan-Kolchin, Bullock, \&
  Kaplinghat}]{10.1111/j.1365-2966.2012.20695.x}
Boylan-Kolchin, M., Bullock, J.~S., \& Kaplinghat, M. 2012, Monthly Notices of
  the Royal Astronomical Society, 422, 1203

\bibitem[{{Bozek} {et~al.}(2015){Bozek}, {Marsh}, {Silk}, \&
  {Wyse}}]{2015MNRAS.450..209B}
{Bozek}, B., {Marsh}, D. J.~E., {Silk}, J., \& {Wyse}, R. F.~G. 2015, \mnras,
  450, 209

\bibitem[{{Brooks} {et~al.}(2013){Brooks}, {Kuhlen}, {Zolotov}, \&
  {Hooper}}]{2013ApJ...765...22B}
{Brooks}, A.~M., {Kuhlen}, M., {Zolotov}, A., \& {Hooper}, D. 2013, \apj, 765,
  22

\bibitem[{{Burkert}(1995)}]{1995ApJ...447L..25B}
{Burkert}, A. 1995, \apjl, 447, L25

\bibitem[{{Burkert}(2015)}]{2015ApJ...808..158B}
{Burkert}, A. 2015, \apj, 808, 158

\bibitem[{{Burkert}(2020)}]{2020arXiv200611111B}
{Burkert}, A. 2020, arXiv e-prints, arXiv:2006.11111

\bibitem[{{Calabrese} \& {Spergel}(2016)}]{2016MNRAS.460.4397C}
{Calabrese}, E. \& {Spergel}, D.~N. 2016, \mnras, 460, 4397

\bibitem[{{de Blok}(2010)}]{2010AdAst2010E...5D}
{de Blok}, W.~J.~G. 2010, Advances in Astronomy, 2010, 789293

\bibitem[{{Di Paolo} \& {Salucci}(2020)}]{2020arXiv200503520D}
{Di Paolo}, C. \& {Salucci}, P. 2020, arXiv e-prints, arXiv:2005.03520

\bibitem[{{Dine} \& {Fischler}(1983)}]{1983PhLB..120..137D}
{Dine}, M. \& {Fischler}, W. 1983, Physics Letters B, 120, 137

\bibitem[{Donato {et~al.}(2009)Donato, Gentile, Salucci, Frigerio~Martins,
  Wilkinson, Gilmore, Grebel, Koch, \& Wyse}]{10.1111/j.1365-2966.2009.15004.x}
Donato, F., Gentile, G., Salucci, P., {et~al.} 2009, Monthly Notices of the
  Royal Astronomical Society, 397, 1169

\bibitem[{{Donato} {et~al.}(2009){Donato}, {Gentile}, {Salucci}, {Frigerio
  Martins}, {Wilkinson}, {Gilmore}, {Grebel}, {Koch}, \&
  {Wyse}}]{2009MNRAS.397.1169D}
{Donato}, F., {Gentile}, G., {Salucci}, P., {et~al.} 2009, \mnras, 397, 1169

\bibitem[{{Du} {et~al.}(2017){Du}, {Behrens}, {Niemeyer}, \&
  {Schwabe}}]{2017PhRvD..95d3519D}
{Du}, X., {Behrens}, C., {Niemeyer}, J.~C., \& {Schwabe}, B. 2017, \prd, 95,
  043519

\bibitem[{{Dubois} {et~al.}(2014){Dubois}, {Pichon}, {Welker}, {Le Borgne},
  {Devriendt}, {Laigle}, {Codis}, {Pogosyan}, {Arnouts}, {Benabed}, {Bertin},
  {Blaizot}, {Bouchet}, {Cardoso}, {Colombi}, {de Lapparent}, {Desjacques},
  {Gavazzi}, {Kassin}, {Kimm}, {McCracken}, {Milliard}, {Peirani}, {Prunet},
  {Rouberol}, {Silk}, {Slyz}, {Sousbie}, {Teyssier}, {Tresse}, {Treyer},
  {Vibert}, \& {Volonteri}}]{2014MNRAS.444.1453D}
{Dubois}, Y., {Pichon}, C., {Welker}, C., {et~al.} 2014, \mnras, 444, 1453

\bibitem[{{Edwards} {et~al.}(2018){Edwards}, {Kendall}, {Hotchkiss}, \&
  {Easther}}]{2018JCAP...10..027E}
{Edwards}, F., {Kendall}, E., {Hotchkiss}, S., \& {Easther}, R. 2018, JCAP,
  2018, 027

\bibitem[{{Flores} \& {Primack}(1994)}]{1994ApJ...427L...1F}
{Flores}, R.~A. \& {Primack}, J.~R. 1994, \apjl, 427, L1

\bibitem[{{Gentile} {et~al.}(2004){Gentile}, {Salucci}, {Klein}, {Vergani}, \&
  {Kalberla}}]{2004MNRAS.351..903G}
{Gentile}, G., {Salucci}, P., {Klein}, U., {Vergani}, D., \& {Kalberla}, P.
  2004, \mnras, 351, 903

\bibitem[{{Gonz{\'a}lez-Morales} {et~al.}(2017){Gonz{\'a}lez-Morales}, {Marsh},
  {Pe{\~n}arrubia}, \& {Ure{\~n}a-L{\'o}pez}}]{2017MNRAS.472.1346G}
{Gonz{\'a}lez-Morales}, A.~X., {Marsh}, D. J.~E., {Pe{\~n}arrubia}, J., \&
  {Ure{\~n}a-L{\'o}pez}, L.~A. 2017, \mnras, 472, 1346

\bibitem[{{Gonz{\'a}lez-Samaniego} {et~al.}(2017){Gonz{\'a}lez-Samaniego},
  {Bullock}, {Boylan-Kolchin}, {Fitts}, {Elbert}, {Hopkins}, {Kere{\v{s}}}, \&
  {Faucher-Gigu{\`e}re}}]{2017MNRAS.472.4786G}
{Gonz{\'a}lez-Samaniego}, A., {Bullock}, J.~S., {Boylan-Kolchin}, M., {et~al.}
  2017, \mnras, 472, 4786

\bibitem[{Governato {et~al.}(2012)Governato, Zolotov, Pontzen, Christensen, Oh,
  Brooks, Quinn, Shen, \& Wadsley}]{10.1111/j.1365-2966.2012.20696.x}
Governato, F., Zolotov, A., Pontzen, A., {et~al.} 2012, Monthly Notices of the
  Royal Astronomical Society, 422, 1231

\bibitem[{{Hlozek} {et~al.}(2015){Hlozek}, {Grin}, {Marsh}, \&
  {Ferreira}}]{2015PhRvD..91j3512H}
{Hlozek}, R., {Grin}, D., {Marsh}, D. J.~E., \& {Ferreira}, P.~G. 2015, \prd,
  91, 103512

\bibitem[{{Hu} {et~al.}(2000){Hu}, {Barkana}, \&
  {Gruzinov}}]{2000PhRvL..85.1158H}
{Hu}, W., {Barkana}, R., \& {Gruzinov}, A. 2000, \prl, 85, 1158

\bibitem[{{Hui} {et~al.}(2017){Hui}, {Ostriker}, {Tremaine}, \&
  {Witten}}]{2017PhRvD..95d3541H}
{Hui}, L., {Ostriker}, J.~P., {Tremaine}, S., \& {Witten}, E. 2017, \prd, 95,
  043541

\bibitem[{{Klypin} {et~al.}(1999){Klypin}, {Kravtsov}, {Valenzuela}, \&
  {Prada}}]{1999ApJ...522...82K}
{Klypin}, A., {Kravtsov}, A.~V., {Valenzuela}, O., \& {Prada}, F. 1999, ApJ,
  522, 82

\bibitem[{Knollmann \& Knebe(2009)}]{Knollmann_2009}
Knollmann, S.~R. \& Knebe, A. 2009, The Astrophysical Journal Supplement
  Series, 182, 608

\bibitem[{{Kopp} {et~al.}(2017){Kopp}, {Vattis}, \&
  {Skordis}}]{2017PhRvD..96l3532K}
{Kopp}, M., {Vattis}, K., \& {Skordis}, C. 2017, \prd, 96, 123532

\bibitem[{{Kormendy}(2015)}]{2015llg..book..323K}
{Kormendy}, J. 2015, {Structure and Evolution of Dwarf Galaxies}, 323

\bibitem[{{Lee} \& {Koh}(1996)}]{1996PhRvD..53.2236L}
{Lee}, J.-W. \& {Koh}, I.-G. 1996, \prd, 53, 2236

\bibitem[{{Li} {et~al.}(2020){Li}, {Lelli}, {McGaugh}, \&
  {Schombert}}]{2020ApJS..247...31L}
{Li}, P., {Lelli}, F., {McGaugh}, S., \& {Schombert}, J. 2020, \apjs, 247, 31

\bibitem[{{Li} {et~al.}(2019){Li}, {Hui}, \& {Bryan}}]{2019PhRvD..99f3509L}
{Li}, X., {Hui}, L., \& {Bryan}, G.~L. 2019, \prd, 99, 063509

\bibitem[{{Madelung}(1926)}]{1926NW.....14.1004M}
{Madelung}, E. 1926, Naturwissenschaften, 14, 1004

\bibitem[{{Marsh}(2015)}]{2015PhRvD..91l3520M}
{Marsh}, D.~J.~E. 2015, \prd, 91, 123520

\bibitem[{{Marsh}(2016)}]{2016PhR...643....1M}
{Marsh}, D.~J.~E. 2016, PhysRep, 643, 1

\bibitem[{{Marsh} \& {Ferreira}(2010)}]{2010PhRvD..82j3528M}
{Marsh}, D. J.~E. \& {Ferreira}, P.~G. 2010, \prd, 82, 103528

\bibitem[{{Marsh} \& {Pop}(2015)}]{2015MNRAS.451.2479M}
{Marsh}, D. J.~E. \& {Pop}, A.-R. 2015, MNRAS, 451, 2479

\bibitem[{{McGaugh} {et~al.}(2000){McGaugh}, {Schombert}, {Bothun}, \& {de
  Blok}}]{2000ApJ...533L..99M}
{McGaugh}, S.~S., {Schombert}, J.~M., {Bothun}, G.~D., \& {de Blok}, W.~J.~G.
  2000, \apjl, 533, L99

\bibitem[{Mina {et~al.}(2019)Mina, Mota, \& Winther}]{mina2019scalar}
Mina, M., Mota, D.~F., \& Winther, H.~A. 2019, SCALAR: an AMR code to simulate
  axion-like dark matter models

\bibitem[{Mocz {et~al.}(2020)Mocz, Fialkov, Vogelsberger, Becerra, Shen,
  Robles, Amin, Zavala, Boylan-Kolchin, Bose, \& et~al.}]{Mocz_2020}
Mocz, P., Fialkov, A., Vogelsberger, M., {et~al.} 2020, Monthly Notices of the
  Royal Astronomical Society, 494, 2027

\bibitem[{{Mocz} {et~al.}(2018){Mocz}, {Lancaster}, {Fialkov}, {Becerra}, \&
  {Chavanis}}]{2018PhRvD..97h3519M}
{Mocz}, P., {Lancaster}, L., {Fialkov}, A., {Becerra}, F., \& {Chavanis}, P.-H.
  2018, \prd, 97, 083519

\bibitem[{{Mocz} {et~al.}(2017){Mocz}, {Vogelsberger}, {Robles}, {Zavala},
  {Boylan-Kolchin}, {Fialkov}, \& {Hernquist}}]{2017MNRAS.471.4559M}
{Mocz}, P., {Vogelsberger}, M., {Robles}, V.~H., {et~al.} 2017, MNRAS, 471,
  4559

\bibitem[{{Moore}(1994)}]{1994Natur.370..629M}
{Moore}, B. 1994, \nat, 370, 629

\bibitem[{{Moore} {et~al.}(1999){Moore}, {Ghigna}, {Governato}, {Lake},
  {Quinn}, {Stadel}, \& {Tozzi}}]{1999ApJ...524L..19M}
{Moore}, B., {Ghigna}, S., {Governato}, F., {et~al.} 1999, ApJ, 524, L19

\bibitem[{Mortlock {et~al.}(2011)Mortlock, Warren, Venemans, Patel, Hewett,
  McMahon, Simpson, Theuns, Gonzales-Solares, Adamson, Dye, Hambly, Hirst,
  Irwin, Kuiper, Lawrence, \& Rottgering}]{f2293ac541c24ba690492c90645c960e}
Mortlock, D., Warren, S., Venemans, B., {et~al.} 2011, Nature, 474, 616

\bibitem[{{Navarro} {et~al.}(1996){Navarro}, {Frenk}, \&
  {White}}]{1996ApJ...462..563N}
{Navarro}, J.~F., {Frenk}, C.~S., \& {White}, S. D.~M. 1996, \apj, 462, 563

\bibitem[{{Nori} \& {Baldi}(2018)}]{2018arXiv180108144N}
{Nori}, M. \& {Baldi}, M. 2018, ArXiv e-prints [\eprint[arXiv]{1801.08144}]

\bibitem[{{Nori} \& {Baldi}(2020)}]{2020arXiv200701316N}
{Nori}, M. \& {Baldi}, M. 2020, arXiv e-prints, arXiv:2007.01316

\bibitem[{Peebles(2000)}]{Peebles_2000}
Peebles, P. J.~E. 2000, The Astrophysical Journal, 534, L127

\bibitem[{{Planck Collaboration} {et~al.}(2018){Planck Collaboration},
  {Aghanim}, {Akrami}, {Ashdown}, {Aumont}, {Baccigalupi}, {Ballardini},
  {Banday}, {Barreiro}, {Bartolo}, {Basak}, {Battye}, {Benabed}, {Bernard},
  {Bersanelli}, {Bielewicz}, {Bock}, {Bond}, {Borrill}, {Bouchet}, {Boulanger},
  {Bucher}, {Burigana}, {Butler}, {Calabrese}, {Cardoso}, {Carron},
  {Challinor}, {Chiang}, {Chluba}, {Colombo}, {Combet}, {Contreras}, {Crill},
  {Cuttaia}, {de Bernardis}, {de Zotti}, {Delabrouille}, {Delouis}, {Di
  Valentino}, {Diego}, {Dor{\'e}}, {Douspis}, {Ducout}, {Dupac}, {Dusini},
  {Efstathiou}, {Elsner}, {En{\ss}lin}, {Eriksen}, {Fantaye}, {Farhang},
  {Fergusson}, {Fernandez-Cobos}, {Finelli}, {Forastieri}, {Frailis},
  {Fraisse}, {Franceschi}, {Frolov}, {Galeotta}, {Galli}, {Ganga},
  {G{\'e}nova-Santos}, {Gerbino}, {Ghosh}, {Gonz{\'a}lez-Nuevo}, {G{\'o}rski},
  {Gratton}, {Gruppuso}, {Gudmundsson}, {Hamann}, {Handley}, {Hansen},
  {Herranz}, {Hildebrandt}, {Hivon}, {Huang}, {Jaffe}, {Jones}, {Karakci},
  {Keih{\"a}nen}, {Keskitalo}, {Kiiveri}, {Kim}, {Kisner}, {Knox},
  {Krachmalnicoff}, {Kunz}, {Kurki-Suonio}, {Lagache}, {Lamarre}, {Lasenby},
  {Lattanzi}, {Lawrence}, {Le Jeune}, {Lemos}, {Lesgourgues}, {Levrier},
  {Lewis}, {Liguori}, {Lilje}, {Lilley}, {Lindholm}, {L{\'o}pez-Caniego},
  {Lubin}, {Ma}, {Mac{\'\i}as-P{\'e}rez}, {Maggio}, {Maino}, {Mandolesi},
  {Mangilli}, {Marcos-Caballero}, {Maris}, {Martin}, {Martinelli},
  {Mart{\'\i}nez-Gonz{\'a}lez}, {Matarrese}, {Mauri}, {McEwen}, {Meinhold},
  {Melchiorri}, {Mennella}, {Migliaccio}, {Millea}, {Mitra},
  {Miville-Desch{\^e}nes}, {Molinari}, {Montier}, {Morgante}, {Moss}, {Natoli},
  {N{\o}rgaard-Nielsen}, {Pagano}, {Paoletti}, {Partridge}, {Patanchon},
  {Peiris}, {Perrotta}, {Pettorino}, {Piacentini}, {Polastri}, {Polenta},
  {Puget}, {Rachen}, {Reinecke}, {Remazeilles}, {Renzi}, {Rocha}, {Rosset},
  {Roudier}, {Rubi{\~n}o-Mart{\'\i}n}, {Ruiz-Granados}, {Salvati}, {Sandri},
  {Savelainen}, {Scott}, {Shellard}, {Sirignano}, {Sirri}, {Spencer},
  {Sunyaev}, {Suur-Uski}, {Tauber}, {Tavagnacco}, {Tenti}, {Toffolatti},
  {Tomasi}, {Trombetti}, {Valenziano}, {Valiviita}, {Van Tent}, {Vibert},
  {Vielva}, {Villa}, {Vittorio}, {Wand elt}, {Wehus}, {White}, {White},
  {Zacchei}, \& {Zonca}}]{2018arXiv180706209P}
{Planck Collaboration}, {Aghanim}, N., {Akrami}, Y., {et~al.} 2018, arXiv
  e-prints, arXiv:1807.06209

\bibitem[{{Preskill} {et~al.}(1983){Preskill}, {Wise}, \&
  {Wilczek}}]{1983PhLB..120..127P}
{Preskill}, J., {Wise}, M.~B., \& {Wilczek}, F. 1983, Physics Letters B, 120,
  127

\bibitem[{{Salucci} \& {Burkert}(2000)}]{2000ApJ...537L...9S}
{Salucci}, P. \& {Burkert}, A. 2000, \apjl, 537, L9

\bibitem[{Salucci {et~al.}(2007)Salucci, Lapi, Tonini, Gentile, Yegorova, \&
  Klein}]{10.1111/j.1365-2966.2007.11696.x}
Salucci, P., Lapi, A., Tonini, C., {et~al.} 2007, Monthly Notices of the Royal
  Astronomical Society, 378, 41

\bibitem[{{Schaye} {et~al.}(2015){Schaye}, {Crain}, {Bower}, {Furlong},
  {Schaller}, {Theuns}, {Dalla Vecchia}, {Frenk}, {McCarthy}, {Helly},
  {Jenkins}, {Rosas-Guevara}, {White}, {Baes}, {Booth}, {Camps}, {Navarro},
  {Qu}, {Rahmati}, {Sawala}, {Thomas}, \& {Trayford}}]{2015MNRAS.446..521S}
{Schaye}, J., {Crain}, R.~A., {Bower}, R.~G., {et~al.} 2015, \mnras, 446, 521

\bibitem[{{Schive} {et~al.}(2014){Schive}, {Chiueh}, \&
  {Broadhurst}}]{2014NatPh..10..496S}
{Schive}, H.-Y., {Chiueh}, T., \& {Broadhurst}, T. 2014, Nature Physics, 10,
  496

\bibitem[{Schive {et~al.}(2014)Schive, Liao, Woo, Wong, Chiueh, Broadhurst, \&
  Hwang}]{Schive_2014}
Schive, H.-Y., Liao, M.-H., Woo, T.-P., {et~al.} 2014, Physical Review Letters,
  113

\bibitem[{{Schwabe} {et~al.}(2016){Schwabe}, {Niemeyer}, \&
  {Engels}}]{2016PhRvD..94d3513S}
{Schwabe}, B., {Niemeyer}, J.~C., \& {Engels}, J.~F. 2016, \prd, 94, 043513

\bibitem[{{Springel}(2005)}]{2005MNRAS.364.1105S}
{Springel}, V. 2005, \mnras, 364, 1105

\bibitem[{{Teyssier}(2002)}]{2002AA...385..337T}
{Teyssier}, R. 2002, AAP, 385, 337

\bibitem[{{Tinker} {et~al.}(2008){Tinker}, {Kravtsov}, {Klypin}, {Abazajian},
  {Warren}, {Yepes}, {Gottl{\"o}ber}, \& {Holz}}]{2008ApJ...688..709T}
{Tinker}, J., {Kravtsov}, A.~V., {Klypin}, A., {et~al.} 2008, \apj, 688, 709

\bibitem[{{Turk} {et~al.}(2011){Turk}, {Smith}, {Oishi}, {Skory}, {Skillman},
  {Abel}, \& {Norman}}]{2011ApJS..192....9T}
{Turk}, M.~J., {Smith}, B.~D., {Oishi}, J.~S., {et~al.} 2011, The Astrophysical
  Journal Supplement Series, 192, 9

\bibitem[{{Uhlemann} {et~al.}(2014){Uhlemann}, {Kopp}, \&
  {Haugg}}]{2014PhRvD..90b3517U}
{Uhlemann}, C., {Kopp}, M., \& {Haugg}, T. 2014, \prd, 90, 023517

\bibitem[{{Vogelsberger} {et~al.}(2014){Vogelsberger}, {Genel}, {Springel},
  {Torrey}, {Sijacki}, {Xu}, {Snyder}, {Bird}, {Nelson}, \&
  {Hernquist}}]{2014Natur.509..177V}
{Vogelsberger}, M., {Genel}, S., {Springel}, V., {et~al.} 2014, \nat, 509, 177

\bibitem[{{Weinberg} {et~al.}(2015){Weinberg}, {Bullock}, {Governato}, {Kuzio
  de Naray}, \& {Peter}}]{2015PNAS..11212249W}
{Weinberg}, D.~H., {Bullock}, J.~S., {Governato}, F., {Kuzio de Naray}, R., \&
  {Peter}, A. H.~G. 2015, Proceedings of the National Academy of Science, 112,
  12249

\bibitem[{{Widrow} \& {Kaiser}(1993)}]{1993ApJ...416L..71W}
{Widrow}, L.~M. \& {Kaiser}, N. 1993, ApJL, 416, L71

\bibitem[{{Woo} \& {Chiueh}(2009)}]{2009ApJ...697..850W}
{Woo}, T.-P. \& {Chiueh}, T. 2009, ApJ, 697, 850

\bibitem[{{Zhang} {et~al.}(2018){Zhang}, {Sming Tsai}, {Kuo}, {Cheung}, \&
  {Chu}}]{2018ApJ...853...51Z}
{Zhang}, J., {Sming Tsai}, Y.-L., {Kuo}, J.-L., {Cheung}, K., \& {Chu}, M.-C.
  2018, ApJ, 853, 51

\end{thebibliography}

\appendix

\section{Initial conditions}\label{sect:initial_conditions}

In order to generate suitable initial conditions, we first obtain the linear matter power-spectrum from the publicly available code \texttt{AxionCAMB}  \citep{2015PhRvD..91j3512H}, which is then used to generate a gaussian realization of $\delta(k,z_{\rm ini})$.

In order to generate the initial wave-function, we consider the Schr\"odinger-Poisson system in the Madelung form. In particular, by expressing the wave-function in polar coordinates:
\begin{align}
\psi = \sqrt{1+\delta} ~ e^{i\theta} ~ .
\end{align}
Again, macroscopic quantities such as density and velocity are defined by means of:
\begin{align}
\delta &= \left | \psi \right |^2 - 1~, \\
\mathbf{v} &= \frac{1}{a m_{\rm B}} \nabla \theta ~.
\end{align}
Here $\delta$ represents an overdensity rather than the density itself. Then, the first Madelung equation can be written as:
\begin{align}
\frac{d\delta}{dt} + \frac{1}{a} \nabla \cdot \left[ \left ( 1+\delta \right ) \mathbf{v} \right ] = 0~. 
\end{align}
In the linear regime, initial conditions can be generated by using the Zel'dovich approximation (see \citet{2017PhRvD..96l3532K} for a more general description):
\begin{align}
\delta(x,z) = \frac{D(z)}{D(z_{\rm ini})} \delta(x,z_{\rm ini}) ~ .
\end{align}
where $D$ is the growth factor. Thus, in the linear regime:
\begin{align}
\nabla^2\theta(x,z_{\rm ini}) = -\frac{m_{\rm B} H(z_{\rm ini})f(z_{\rm ini})}{(1+z_{\rm ini})^2} \delta(x,z_{\rm ini}) ~ .
\end{align}
where $f \equiv d\log D/d\log a$ is the growth rate. The Fourier transform of the phase is then calculated according to:
\begin{align}
\theta(k,z_{\rm ini}) = \frac{m_{\rm B} H(z_{\rm ini})f(z_{\rm ini})}{(1+z_{\rm ini})^2} \frac{\delta(k,z_{\rm ini})}{k^2}~.
\end{align}
Then, overdensity and phase in real space are obtained by performing a backward Fourier transform.

\end{document}